\documentclass[nohyper]{JHEP3}

\usepackage[dvips]{graphicx,epsfig}
\usepackage{amsmath}
\usepackage{amsfonts}   
\usepackage{amssymb}  
\usepackage{mathrsfs}
\usepackage{exscale}

\def\be{\begin{equation}}
\def\ee{\end{equation}}
\def\bea{\begin{eqnarray}}
\def\eea{\end{eqnarray}}

\def\fun#1#2{\lower3.6pt\vbox{\baselineskip0pt\lineskip.9pt
        \ialign{$\mathsurround=0pt#1\hfill##\hfil$\crcr#2\crcr\sim\crcr}}}
        
\def\expec#1{\langle#1\rangle}

\def\vmu{\mbox{\boldmath${\mu}$}}
\def\bfp{\mbox{\bf p}}
\def\bfq{\mbox{\bf q}}

\def\x{{\bf x}}
\def\C{{\bf C}}

\newcommand{\DA}{D\!_A(z)}
\newcommand{\hz}{H(z)}

\newcommand{\Vsur}{V_{\rm survey}}
\newcommand{\Veff}{V_{\rm eff}}

\newcommand{\w}{w}

\newcommand{\ihMpc}{h{\rm\;Mpc^{-1}}}
\newcommand{\kmax}{k_{\rm max}}

\title{Measuring the neutrino mass from future wide galaxy cluster catalogues}

\author{Carmelita Carbone\\  
Dipartimento di Astronomia, Alma Mater Studiorum-Universit\`a di
Bologna, via Ranzani 1, I-40127 Bologna, Italy \\ 
\& 
INAF-Osservatorio Astronomico di Bologna, Via Ranzani 1, I-40127,
Bologna, Italy\\
\&
INFN, Sezione di Bologna, Viale Berti Pichat 6/2, I-40127 Bologna, Italy\\
\email{carmelita.carbone@unibo.it}}
\author{Cosimo Fedeli\\ 
Department of Astronomy, University of Florida, 211 Bryant Space
Science Center, Gainesville, FL 32611\\
\email{cosimo.fedeli@astro.ufl.edu}}
\author{Lauro Moscardini\\
Dipartimento di Astronomia, Alma Mater Studiorum-Universit\`a di
Bologna, via Ranzani 1, I-40127 Bologna, Italy \\
\&
INAF-Osservatorio Astronomico di Bologna, Via Ranzani 1, I-40127,
Bologna, Italy\\
\&
INFN, Sezione di Bologna, Viale Berti Pichat 6/2, I-40127 Bologna, Italy\\
\email{lauro.moscardini@unibo.it}}
\author{Andrea Cimatti\\
Dipartimento di Astronomia, Alma Mater Studiorum-Universit\`a di
Bologna, via Ranzani 1, I-40127 Bologna, Italy \\
\& INAF-Osservatorio Astronomico di Bologna, Via Ranzani 1, I-40127,
Bologna, Italy\\   
\email{a.cimatti@unibo.it}}

\abstract{We present forecast errors on a wide range of cosmological
  parameters obtained from a photometric cluster catalogue of a future
  wide-field \emph{Euclid}-like survey. We focus in particular on the
  total neutrino mass as constrained by a combination of the galaxy
  cluster number counts and correlation function. For the
  latter we consider only the shape information and the Baryon
  Acoustic Oscillations (BAO), while
  marginalising over the spectral amplitude and the redshift space
  distortions. In addition to the cosmological parameters of the
  standard $\Lambda$CDM$+\nu$ model we also consider a non-vanishing
  curvature, and two parameters describing a redshift evolution for
  the dark energy equation of state. For completeness, we also
  marginalise over a set of ``nuisance'' parameters, representing the
  uncertainties on the cluster mass determination. We find that
  combining cluster counts with power spectrum information greatly
  improves the constraining power of each probe taken individually,
  with errors on cosmological parameters being reduced by up to an
  order of magnitude. In particular, the best improvements are for the parameters defining
the dynamical evolution of dark energy, where cluster counts break
  degeneracies.
  Moreover, the resulting error on neutrino mass is at the
  level of $\sigma(M_\nu)\sim 0.9$ eV, comparable with that derived from
  present Ly$\alpha$ forest measurements and Cosmic Microwave
  background (CMB) data in the
  framework of a non-flat Universe. Further adopting \emph{Planck} priors
  and reducing the number of free parameters to a $\Lambda$CDM+$\nu$
  cosmology allows to place
  constraints on the total neutrino mass of $\sigma(M_\nu) \sim 0.08$
  eV, close to the lower bound enforced by
  neutrino oscillation experiments. Finally, in the optimistic case
  where uncertainties in the calibration of the mass-observable
  relation were so small to be neglected, the combination of \emph{Planck}
  priors with cluster counts and power spectrum would constrain the
  total neutrino mass down to $\sigma(M_\nu) \sim 0.034$ eV, i.e. the
  minimum neutrino mass predicted by oscillation experiments would be
  detected in a $\Lambda$CDM framework.
  We thus show that galaxy clusters
  from future wide galaxy surveys will be an excellent tool for
  studying cosmology and fundamental physics.}

\begin{document}

\section{Introduction}
\label{Intro}
It is now established from solar, atmospheric, reactor
and accelerator experiments that neutrinos have non-zero mass,
and that a lower limit on the total neutrino mass is given by
$M_\nu\equiv \sum m_{\nu}\sim 0.05$ eV \cite{lesgourgues2006, Wong2011}, where
$m_\nu$ is the mass of a single neutrino species. On the other hand
the individual masses are still unknown. Since neutrino mass affects the
evolution of the Universe in several observable ways, its measurements
can be obtained from different cosmological probes, such as observations of
the CMB, galaxy clustering, Ly$\alpha$
forest, and weak lensing data \cite{abazajian2011,Joudaki&Kaplinghat2011,Joudaki2012,Signe_etal2011}.

In particular, a thermal neutrino relic component in the Universe
impacts both the expansion history and the growth of cosmic
structures. Neutrinos with mass $\lesssim 0.6$ eV become
non-relativistic after the epoch of recombination probed by the CMB,
and this mechanism allows massive neutrinos to alter the
matter-radiation equality for a fixed $\Omega_m h^2$. 
Massive neutrinos act as non-relativistic
particles on scales $k>k_{\rm nr}=0.018(m_\nu/1{\rm
  eV})^{1/2}\Omega_m^{1/2} h$/Mpc, where $k_{\rm nr}$ is the
wave-number corresponding to the Hubble horizon size at the epoch
$z_{\rm nr}$ when the given neutrino species becomes non-relativistic,
$\Omega_m$ is the matter energy density and $h=H_0/100\, {\rm km\,
  s^{-1} Mpc^{-1}}$. The large velocity dispersion of non-relativistic
neutrinos suppresses the formation of neutrino perturbations in a way
that depends on $m_\nu$ and redshift $z$, leaving an imprint on the
matter power spectrum for scales $k>k_{\rm fs}(z)=0.82
H(z)/H_0/(1+z)^2 (m_\nu/1{\rm eV}) h$/Mpc
\cite{lesgourgues2006,takada2006}, where neutrinos cannot cluster and
do not contribute to the gravitational potential wells produced by
cold dark matter and baryons.  This modifies the shape of the matter
power spectrum and the correlation function on these scales 
\cite{doroshkevich1981,hu1998,abazajian2005,kiakotou2008,brandbyge2010,Matteo2,Takada2011}.

Massive neutrinos affect also the CMB statistics. WMAP7 alone
constrains $M_\nu < 1.3$ eV \cite{Komatsuetal2010} and data from the
ACT\footnote{http://wwwphy.princeton.edu/act/} and
SPT\footnote{http://pole.uchicago.edu/} experiments constrain
$M_\nu < 0.948$ eV in the framework 
of a non-flat cosmology \cite{Smith_etal2011}. Furthermore, thanks to the
improved sensitivity to polarisation and to the angular power spectrum
damping tail, forecasts for the \emph{Planck} satellite alone give a
1--$\sigma$ error on the total neutrino mass of $\sim 0.2-0.4$ eV,
depending on the assumed cosmological model and fiducial neutrino mass
\cite{perotto2006, kitching2008}.
Moreover, the combination of present data-sets from CMB and
large-scale structure (LSS) yields an upper limit of $M_\nu<0.3$ eV
\cite{wang2005,vikhlinin2009,thomas2010,gonzalez2010,reid2010}.
A further robust constraint on neutrino masses has been obtained using
the Sloan Digital Sky Survey flux power spectrum alone, finding an
upper limit of $M_\nu<0.9$ eV ($2\sigma$ C. L.) \cite{Matteo1}.
However, the tightest constraints to date in terms of
a 2$\sigma$ upper limit on the neutrino masses have been obtained by
combining the Sloan Digital Sky Survey flux power from the Ly$\alpha$ 
forest with CMB and galaxy clustering data. These constraints are
  partly driven by a discrepancy in the measured $\sigma_8$ between
  Ly$\alpha$ forest and CMB, and result in $\Sigma m_\nu<0.17$ eV
\cite{seljak2006}. 
Somewhat less constraining bounds
have been obtained by \cite{goobar2006}, while for forecasts on
future CMB and Ly$\alpha$ forest joint constraints we refer to
\cite{gratton2008}.

Moreover, the forecast sensitivity of future LSS experiments, when combined
with \emph{Planck} CMB priors, indicates that observations should soon be
able to detect signatures of the cosmic neutrino background and
measure the neutrino mass even in the case of the minimum mass
$M_\nu=0.05$ eV
\cite{kitching2008,hannestad2007,lsst2009,hannestad2010,lahav2010, Melita11}.
Furthermore, these future surveys have been planned
to measure with high accuracy the so-called ``dark energy''
equation of state. In fact, there is now strong evidence that the current energy density of the
Universe is dominated by dark energy with an equation of state $w \sim
-1$, which is causing accelerated expansion. Accordingly, in this work
we consider, in addition to massive neutrinos, the effect on LSS of a homogeneous dark energy component
with a general time-varying equation of state $w(z)$.

In this respect, Ref.~\cite{Melita11} has shown that future spectroscopic
galaxy surveys, such as the ESA selected mission
\emph{Euclid}\footnote{http://www.euclid-ec.org/} \cite{RedBook},
will be capable of estimating the neutrino mass scale independently
of flatness assumptions and dark energy parametrisation, if the total neutrino
mass $M_\nu$ is $>0.1$ eV. On the other hand, if $M_\nu$ is $<0.1$ eV,
the sum of neutrino masses, and in particular the minimum neutrino
mass required by neutrino oscillations, can be measured in the context
of a $\Lambda$CDM model. For further discussion on neutrino mass
constraints from different probes see e.g. \cite{Wong2011,abazajian2011} and
references therein.

In this picture, little attention has been dedicated to the
constraining potential of galaxy clusters. Historically these systems
have played a fundamental role in cosmology. For instance, the
correlation function of galaxy clusters provided the very first hint
toward the low-density Universe that is today commonly accepted as the
standard cosmological model \cite{Bahcall1,Bahcall2}. Galaxy cluster
number counts at high-redshift later provided yet another early
evidence for the matter density parameter being substantially smaller
than unity \cite{Carlberg,Donahue}. Galaxy clusters trace the
large-scale matter distribution much as galaxies, with the difference
that the former have a substantially larger bias than the latter, and
hence their correlation is a factor of a few higher. At the
same time, clusters are relatively rare objects, therefore their
abundance is highly sensitive to features in the primordial matter
power spectrum. These occurrences make clusters in principle valuable
tools for constraining the details of the cosmological model, and in
particular the neutrino masses. However their low spatial number
density is also a drawback, in that it increases substantially the
shot noise and the Poisson noise as compared to galaxies. Their
effective cosmological power hence depends significantly on the chosen
selection function and on the precision with which cluster masses can
be estimated. In this work we study how the combination of cluster
number counts with the shape of the cluster power spectrum, including
information from the BAO, can
constrain the total neutrino mass. We focus on the cluster catalogue
that will be produced by the photometric part of \emph{Euclid}, and
provide constraints on the parameter set of an extended cosmological
model which includes curvature, massive neutrinos, and a time-varying
dark energy component. 
Since deviations from the standard model in the form of
extra neutrino species are still uncertain
\cite{giusarma2011,gonzalez-morales2011}, in this paper we focus on
standard neutrino 
families only and analyse the constraining power that future galaxy
cluster catalogues, photometrically selected from galaxy surveys of
\emph{Euclid}-type, have on the total neutrino mass.

The rest of the paper is organised as
follows. In \S~\ref{Clusters} we present the galaxy cluster modelling adopted in this
work and the modifications due to uncertainties in the calibration of the mass-observable
relation. In \S~\ref{Fiducial cosmologies} we describe the adopted fiducial model and the
effect of massive neutrinos on the matter power spectrum. In \S~\ref{Fishers}
we review our forecasting approach as applied to cluster counts and
power spectrum, and specify characteristics of the galaxy cluster
survey analysed in this work. In \S~\ref{Results} we present our results on
the forecast neutrino mass errors, and finally in \S~\ref{Conclu} we draw
our conclusions.

\section{Clusters: effective bias and mass function}
\label{Clusters}

In the cosmological analysis performed in this paper we
combine number counts and the power spectrum of
galaxy clusters. These are identified as overdensities of
  galaxies photometrically selected with a \emph{Euclid}-type survey.
In order to properly define the
relevant cluster catalogue we adopt the minimum mass
$M_{\mathrm{min}}(z)$ provided by the \emph{Euclid Red-Book} \cite{RedBook}  for
objects identified as having a S/N ratio larger than $5$. Let us
assume a perfect knowledge of the mass of each cluster in the catalogue
for the time being. This assumption will be relaxed in
\S~\ref{Nuisance} below. The resulting (full-sky equivalent) cluster redshift distribution
can then be computed as
\begin{equation}
\frac{dN(z)}{dz} = \frac{dV(z)}{dz}~g(z)~,
\end{equation}
where $dV(z)/dz$ is the cosmic volume per unit redshift, while
\begin{equation}
\label{eqn:gz}
g(z)\equiv \int_{M_{\mathrm{min}}(z)}^{+\infty} dM~n(M,z)~.
\end{equation}
For the mass function $n(M,z)$ of dark matter halos we adopt the
prescription given by \cite{Sheth&Tormen2002}, which is based on an 
approximated ellipsoidal collapse model and has been shown to agree 
well with the results of numerical cosmological simulations. 
Although derived in a $\Lambda$CDM context, this prescription has
been shown to model $n(M,z)$ accurately even in the presence of
dynamic dark energy \cite{Percival05,Wang&Steinhardt} and massive
neutrinos \cite{Marulli_etal2011a,Takada2011,Jose_etal2011},
and holds, in particular, for the fiducial values of neutrino mass
and dark energy parameters considered in this work (see \S
\ref{Fiducial cosmologies}).

As for the power spectrum of galaxy clusters, we simply assume it to
be a biased version of the linear dark matter power
spectrum. Nonlinear effects become important on scales $k \gtrsim
0.1~h$ Mpc$^{-1}$, which at low redshift have been disregarded in our
Fisher matrix analysis, as we shall explain in Appendix \ref{appendix:Pobs}. By
assuming again a perfect knowledge of the cluster masses in the
\emph{Euclid} catalogue, we can compute the power spectrum as $P(k,z) =
b_{\mathrm{e}}^2(z)~P_{\mathrm{L}}(k,z)$, where $P_{\mathrm{L}}(k,z)$ is the
linear dark matter power spectrum, while $b_{\mathrm{e}}(z)$ is the
effective bias of clusters in the catalogue, defined as 
\begin{equation}
\label{eqn:eb}
b_{\mathrm{e}}(z) \equiv \frac{1}{g(z)}
\int_{M_{\mathrm{min}}(z)}^{+\infty} dM~n(M,z)~b(M,z)~.
\end{equation}
In Eq.~(\ref{eqn:eb}), the function $b(M,z)$ represents the bias of
dark matter halos, for which we adopt the semi-analytic prescription
of \cite{Sheth_Mo_Tormen2001} (see \cite{Marulli_etal2011a} for an
analysis of this prescription against N-body simulations which include a
massive neutrino component). 
We compute the linear dark matter
power spectrum $P_\mathrm{L}(k,z)$ with the publicly available software package
CAMB \cite{CAMB}, which takes correctly into account the effect of
massive neutrinos. 

\subsection{Nuisance parameters}
\label{Nuisance}
The discussion presented in \S~\ref{Clusters} assumes a perfect
knowledge of the true mass of clusters in our catalogue. This is a very
strong assumption, in that reliable mass estimates can be obtained
only with an extensive multi-wavelength follow-up and only for the
most massive objects. A less expensive alternative,
particularly suitable for large cluster catalogues, is to adopt a
scaling relation between the observable quantity at hand (cluster
richness in this case) and the true underlying mass. Scaling 
relations between cluster properties are however not perfect
one-to-one associations; they include a scatter and, in some
circumstances, systematic biases. In order to take properly into
account the uncertainties that a scaling relation introduces in the
knowledge of the true underlying mass, we treat the scatter and
systematic biases as ``nuisance'' parameters, and marginalise over them
in the Fisher matrix analysis for cluster counts in \S~\ref{Counts}.

Given this, let $p(M_{\mathrm{o}}|M)$ be the probability that for a
given cluster of intrinsic mass $M$ we infer the mass $M_{\mathrm{o}}$
through the scaling relation \cite{Lima&Hu2005}. The
redshift distribution of clusters given by Eq.~(\ref{eqn:gz}) above
then gets modified according to  
\begin{equation}
g(z) = \int_{M_{\mathrm{min}}(z)}^{+\infty}
dM_{\mathrm{o}}\int_0^{+\infty}dM~n(M,z)~p(M_{\mathrm{o}}|M). 
\end{equation}
By assuming a lognormal scatter around the nominal scaling relation
with dispersion $\sigma_{\ln M}$, the probability can be written as 
\begin{equation}
p(M_{\mathrm{o}}|M) = \frac{1}{M_{\mathrm{o}}\sqrt{2\pi}\sigma_{\ln M}}
\exp\left[ -x^2(M_{\mathrm{o}}) \right], 
\end{equation}
where 
\begin{equation}
x(M_{\mathrm{o}}) \equiv \frac{1}{\sqrt{2}\sigma_{\ln M}}\left[ \ln
  (M_{\mathrm{o}}) -B_M - \ln (M)\right]. 
\end{equation}
Evidently, the parameter $B_M$ represents the fractional value of a
systematic bias in the scaling relation. It easily follows that the
function $g(z)$ can then be simplified to 
\begin{equation}
g(z) = \frac{1}{2} \int_0^{+\infty}
dM~n(M,z)~\mathrm{erfc}\left[x(M_{\mathrm{min}}(z))\right], 
\label{g_final}
\end{equation}
where erfc$(x)$ is the complementary error function \cite{Abramowitz&Stegun72}.

Likewise, the effective bias presented in Eq.~(\ref{eqn:eb}) now takes
the more general form 
\begin{equation}
b_{\mathrm{e}}(z) = \frac{1}{2g(z)}\int_0^{+\infty}
dM~n(M,z)~b(M,z)~\mathrm{erfc}\left[x(M_{\mathrm{min}}(z))\right]. 
\end{equation}
It is straightforward to verify that, in the ideal case in which the
systematic bias $B_M \rightarrow 0$ and the scatter
$\sigma_{\ln M}\rightarrow 0$, the expressions for the cluster redshift
distribution and the effective bias reduce to their simplest form,
presented at the beginning of \S \ref{Clusters}. In what follows
we adopt these updated forms for 
the cluster redshift distribution and the effective bias.

Moreover, following \cite{Lima&Hu2005}, we assume
the following redshift parametrisation for the halo mass bias and variance:
\begin{eqnarray}
\ln B_M(z) & = & A+B \ln(1+z) \nonumber \\
\sigma_{\ln M}^2(z) & = & \sigma_{\ln M,0}^2-1+(1+z)^{2\beta} \,.
\label{eq:nuis}
\end{eqnarray}
In this way, we have four nuisance parameters, namely $A$, $B$, $\sigma_{\ln
  M,0}$ and $\beta$. For the reference model, we assume $\ln
B_M(z)$  to be zero at $z=0$ (i.e. $A=0$), with no evolution
(i.e. $B=0$), but leaving $A$ and $B$ as free parameters with Gaussian
priors of $\sigma(A)=\sigma(B)=0.25$ \cite{Johnston_etal2007},
which is not overly restrictive for \emph{Euclid} \cite{RedBook}. In
Eq.~(\ref{eq:nuis}), we have
also allowed for a power-law redshift evolution for the variance $\sigma_{\ln M}^2(z)$ of the lognormal intrinsic scatter around the nominal scaling relation. Ref.~\cite{Rykoff_etal2011} estimates
that $\sigma_{\ln M,0}=0.2$ and we choose $\beta=0.125$. This means
that the scatter will grow to a value of $0.6$ at a redshift of
$z=2$. In the count Fisher matrix analysis we self-calibrate for these
scatter variables 
with Gaussian priors of $\sigma(\sigma_{\ln M,0})=0.1$
\cite{Tinker_etal2011} and $\sigma(\beta)=0.1$, which should be
conservative estimates, although these quantities are 
unconstrained by present data
\cite{RedBook}. It is expected that Stage III surveys,
e.g. DES\footnote{http://www.darkenergysurvey.org/}, will shed light
on this quantity beyond redshift $z=1$.

As mentioned above, for the minimum cluster mass as a function of redshift
$M_\mathrm{min}(z)$, we adopt the prescription detailed in the
\emph{Euclid Red-Book} \cite{RedBook} for the photometric cluster
catalogue. The redshift evolution of the cluster number counts is
computed in bins of width $\Delta z=0.1$ between $z=0.2$ and $z=2$,
and we integrate the cluster distribution over mass, considering
number counts in redshift-space only. Concerning redshift errors
that enter the observed galaxy cluster power spectrum, better described in
Appendix \ref{appendix:Pobs}, we assume that photometrically
selected clusters will be spectroscopically confirmed up to $z=1$. 

\section{Cosmological model with massive neutrinos}
\label{Fiducial cosmologies}
According to the latest observations (e.g. \cite{Komatsuetal2010,Larson11} and
refs. therein), we assume the following fiducial cosmological model at
the present epoch:
$\Omega_m=0.271$, $\Omega_\Lambda=1-\Omega_m$, $h=0.703$,  
$A_s=2.525\times 10^{-9}$, $\Omega_b=0.045$, $n_s=0.966$, $w_0=-0.95$,
$w_a=0$, $M_\nu \equiv \sum_i m_{\nu_i}=0.05 {\rm eV}${}\footnote{Here
  we assume an 
  effective number of neutrino species 
  $N_{\rm eff}=3.04$ and, since neutrino oscillation experiments
  have shown that at least one neutrino species is heavier than $0.05$
  eV, we consider an inverted hierarchy  where $m_1 \sim m_2,
  m_3 \sim 0$}. This corresponds to $\sigma_8=0.8$ for the fiducial cosmology.
We consider neither primordial gravitational waves nor a scale
dependent component of the scalar spectral index, and 
assume the matter energy density $\Omega_{m}$ to include the baryon
and neutrino contributions when neutrinos are non-relativistic, so that
$\Omega_{m} = \Omega_{c} + \Omega_{b} + \Omega_{\nu}$, where
$\Omega_{\nu}=M_\nu/(93.14 h^2 {\rm eV})$ \cite{Mangano_etal2005}.

In our fiducial cosmology, the dark energy is described by a cosmic
fluid with an equation of state $\w_{de}(z)=p_{de}(z)/\rho_{de}(z)$,
where $p_{de}$ and $\rho_{de}$ are the pressure and
energy density of the dark energy fluid, respectively. The redshift dependent 
dark energy density is then
\begin{equation}
\label{eq:rhox}
\rho_{de}(z)=\rho_{de}(0) \exp \left[ 3 \int_0^z
  \frac{1+\w(z')}{1+z'}dz'\right]
\end{equation}
which we normalise so that $\Omega_{de}=\Omega_\Lambda$ at the
present epoch.
Finally, to compute our forecasts on dark energy parameters, we adopt
the widely used linear dark energy equation of state $w_{de}(a)=w_0+(1-a)w_a$
\cite{Chev01,Linder03}, where $a\equiv 1/(1+z)$ is the scale factor
normalised to unity at present.
When the dark energy equation of state is a function of redshift, as
we assume in the present work, the constraints on the the sum of neutrino
masses can degrade significantly, since dark energy and massive
neutrinos both affect the growth rate of 
structures \cite{Saito1}. However, as we will show in \S~\ref{Results},
the combination of CMB and galaxy cluster data reduces or even breaks these
degeneracies. 

As we have mentioned in \S~\ref{Intro}, massive neutrinos suppress the
matter power spectrum on wave-numbers above the free-streaming scale
$k_{\rm fs}$, while on very large scales they behave as ordinary cold
dark matter. The suppression effect is encapsulated in the parameter
$f_\nu \equiv \Omega_\nu/\Omega_m$, which damps the source of matter
density perturbations. In fact, the linear growth rate is defined as
$f_g\equiv d\ln \delta/d \ln a$, where $\delta$ 
represents the matter density perturbation $\delta \equiv
\delta\rho_{m}/\rho_{m}$, and $\rho_{m}$  
and $\delta \rho_{m}$ the matter density and the overdensity,
respectively. In the presence of massive neutrinos and a dark energy
component the equation for the linear evolution of matter
density perturbations on scales $k\gg k_{\rm fs}$ can be written as
\cite{0709.0253,Percival05} 
\begin{equation}
\label{eq:second}
\frac{df_g}{d\ln a}=-f_g^2-\left\{1-\frac{1}{2}[\Omega_{m}(a)+(3w_{de}(a)+1)\Omega_{de}(a)]\right\}f_g+\frac{3}{2}\Omega_{m}(a)(1-f_\nu),
\end{equation}
where $\Omega_{m}(a)=H_{0}^{2}
\Omega_{m}a^{-3}/H^{2}(a)$ 
and $\Omega_{de}(a)=H_{0}^{2} \Omega_\Lambda X(a)/H^{2}(a)$ are
the time-dependent density parameters of matter and dark energy,
respectively. Here $H(a)=H_0 (\Omega_{rad}^{-4}+\Omega_m
a^{-3}+\Omega_K a^{-2}+\Omega_\Lambda X(a))^{1/2}$, where, for the
parametrisation of the dark energy 
equation of state chosen above, we have $X(a)=a^{-3(1+w_0+w_a)}
{\exp}(3w_a(a-1))$ \cite{{Wang&Tegmark04}}.

Eq.~(\ref{eq:second}) is a simplified description of the effect 
of massive neutrinos on the growth of structures, since in the
presence of massive neutrinos the growth rate is not only
redshift-dependent, but also \emph{scale}-dependent. For this reason,
in the present work we have computed $f_g$ using the semi-analytic
formula of Ref.~\cite{0709.0253}, as we explain in Appendix \ref{appendix:Pobs}.

\section{Forecast approach: combining cluster BAO and counts}
\label{Fishers}
In this work we derive neutrino constraints combining measurements from
a \emph{Euclid}-like galaxy cluster catalogue with CMB measurements as obtained from a \emph{Planck}-like CMB
experiment\footnote{www.rssd.esa.int/index.php?project=planck}.  
To this aim, we adopt a Fisher matrix approach \cite{Fisher} that
allows us to forecast cosmological parameter errors from LSS and CMB. 
In order to compute CMB priors we use the specifications of the 
\emph{Planck} satellite and, as
explained in Appendix \ref{sec:Planck}, we describe CMB 
temperature and polarisation power spectra using the parameter set
$\boldsymbol{\theta}= \{\omega_m, \omega_b, \omega_\nu, 
100\theta_S,\log (10^{10} A_s), n_S, \tau\}$, 
where $\theta_S$ is the angular size of the sound horizon at last scattering, and
$\tau$ is the optical depth due to reionisation. After marginalising
over the optical depth, we propagate the \emph{Planck} CMB Fisher matrix $F_{ij}^{\rm CMB}$
into the final sets of parameters ${\bf q}$ adopted in \S~\ref{Pk-method},
using the appropriate Jacobian for the involved parameter
transformation.

\subsection{Fisher matrix for BAO+$\rm{\bf P_{c}(k)}$-shape}
\label{Pk-method}
In this section we apply, to the observed galaxy cluster power spectrum,
the so-called ``$P(k)$ method marginalised over growth information'', which
exploits only the shape and the BAO positions of the cluster power spectrum,
while marginalising over amplitude and redshift-space
distortions. This method allows us to estimate, from the galaxy 
cluster catalogue, measurements of the cosmological 
parameters which characterise the underlying fiducial cosmology. Up to
now, this method has been applied to the galaxy power spectrum in a
number of works (see e.g.,
\cite{Melita11,SE03,Wang06,Wang08,Wang_etal2010,Carbone_etal2011,Amendola_etal2011,diporto1,diporto2}).
Here we present the main formulae describing this approach, referring
to Appendix \ref{appendix:Pobs} for details.

Including redshift-space distortions and the geometrical effects due
to the incorrect assumption of the reference cosmology with respect to
the true one \cite{SE03}, the observed galaxy cluster power spectrum can be
written as
\begin{align}
P_{\rm obs}(k_{{\rm ref}\perp},k_{{\rm ref}\parallel},z)
=\frac {\DA _{\rm ref} ^2 \hz}{\DA ^2 \hz _{\rm ref}} P_{\rm
  c}(k_{{\rm ref}\perp},k_{{\rm ref}\parallel},z) +P_{\rm shot}\,,
\label{eq:Pobs_here}
\end{align}
where $P_{\rm c}$ is given by Eq.~(\ref{eq:Pg}) and $\DA$, $\hz$,
$k_{\perp}$, $k_{\parallel}$ are defined in Appendix \ref{appendix:Pobs}.

We divide the survey volume into redshift shells with size $\Delta
z=0.1$, centred at redshift $z_i$, and choose the following set of
parameters to describe $P_{\rm obs}(k_{{\rm ref}\perp},k_{{\rm ref}\parallel},z)$: 
\begin{equation}
\left\{H(z_i), D_A(z_i), \bar{G}(z_i), \beta(z_i,k), 
P_{shot}^i, \omega_m, \omega_b, \omega_\nu, n_s, h\right\}, 
\end{equation}
where $\omega_\nu\equiv\Omega_\nu h^2$, 
$\omega_m=\Omega_m h^2$, $\omega_b=\Omega_b h^2$.
Finally,  since the growth factor $G(z)$, the effective bias
$b_{\mathrm{e}}(z)$, and the power spectrum 
normalisation $P_0$ are completely degenerate, we introduce the
quantity $\bar{G}(z_i)=(P_0)^{0.5} b_{\mathrm{e}}(z_i) G(z_i)/G(z_0)$ \cite{Wang08}.

Under the assumptions discussed in Appendix \ref{appendix:Pobs}, the
Fisher matrix associated to the observed galaxy cluster power 
spectrum can be approximated as
\cite{Tegmark,Tegmark97} 
\begin{eqnarray}
F_{ij}^{\rm P_c}&=&\int_{\vec{k}_{\rm min}} ^ {\vec{k}_{\rm max}} \frac{\partial
  \ln P_{\rm obs}(\vec{k})}{\partial p_i} \frac{\partial \ln P_{\rm
    obs}(\vec{k})}{\partial p_j} \Veff(\vec{k})
\frac{d\vec{k}}{2(2 \pi)^3}\\ \nonumber
&=&\int_{-1}^{1} \int_{k_{\rm min}}^{\kmax}\frac{\partial \ln
  P_{\rm obs}(k,\mu)}{\partial p_i} \frac{\partial \ln P_{\rm obs}(k,\mu)}{\partial p_j} 
\Veff(k,\mu) \frac{2\pi k^2 dk d\mu}{2(2\pi)^3},
\label{Fisher}                 
\end{eqnarray}
where the derivatives are evaluated at the parameter values $p_i$
of the fiducial model,
and $\Veff$ is the effective volume of the survey:
\begin{eqnarray}
\Veff(k,\mu) =
\left [ \frac{{n_{\rm c}}P_{\rm c}(k,\mu)}{{n_{\rm c}}P_{\rm
      c}(k,\mu)+1} \right ]^2 \Vsur. 
\label{V_eff} 
\end{eqnarray}
Here we have assumed that the comoving galaxy cluster number density
$n_{\rm c}$ is constant in position and given by Eq.~(\ref{g_final}).
Due to azimuthal symmetry around the line of sight,
the three-dimensional galaxy cluster power spectrum
$P_{\rm obs}(\vec{k})$ depends only on $k$ and $\mu$, i.e. it is reduced
to two dimensions by symmetry \cite{SE03}.

We do not include information from the amplitude $\bar{G}(z_i)$ and
the redshift space distortions $\beta(z_i,k)$, so we marginalise over
these parameters and also over $P_\mathrm{shot}^i$. Then we
project $\bfp=\{H(z_i), D_A(z_i), \omega_m, \omega_b, \omega_\nu, n_s, h\}$
into the final sets $\bfq$ of cosmological parameters \cite{Wang08a},
\begin{equation}
\bfq=\left\{\Omega_m,\Omega_{K},\Omega_b, h, M_\nu, n_s, w_0, w_a,
  \log(10^{10} A_s)\right\}.
\label{q_set}
\end{equation}

The transformation from one set of parameters to another is given by
\begin{equation}
F_{\alpha \beta}^{\rm P_c}= \sum_{ij} \frac{\partial p_i}{\partial q_{\alpha}}\,
F_{ij}^{\rm P_c}\, \frac{\partial p_j}{\partial q_{\beta}},
\label{eq:Fisherconv}
\end{equation}
where $F_{\alpha \beta}^{\rm P_c}$ is the survey 
Fisher matrix for the set of parameters $\bfq$, and 
$F_{ij}^{\rm P_c}$ is the survey Fisher matrix for the set of equivalent 
parameters $\bfp$.

The adopted full $P(k)$ method marginalised over
 growth information does not give any constraint on $A_s$, 
since the normalisation of the cluster power spectrum is marginalised
over. Therefore, the $\log(10^{10} A_s)$--errors  
shown in \S~\ref{Results} come from combining cluster counts and CMB
measurements. 

\subsection{Fisher matrix for cluster counts}
\label{Counts}
Following the approach of \cite{holder01,majumdar03,sartoris1,sartoris2,Shimon_etal2010},
the Fisher matrix for the number of clusters, $N_{i}$, within the
$i$-th redshift bin and mass $M>M_{\rm min}(z)$, can be written as
\begin{equation}
  F^{\rm N}_{\alpha \beta}= \sum_{i} \frac{\partial N_{i}}{\partial
    \tilde{q}_\alpha}\frac{\partial N_{i}}{\partial \tilde{q}_\beta}
  \frac{1}{N_{i}}\,, 
\label{eq:fm_nc}
\end{equation}
where the sum over $i$ runs over redshift intervals and the number of 
clusters expected in a survey having a sky 
coverage $\Delta\Omega$ in the redshift range between $z_i$ and $z_{i+1}$
can be written as
\begin{eqnarray}
N_{i} & = & \Delta\Omega \int_{z_i}^{z_{i+1}}dz\,{dV\over dz
  d\Omega} g(z).
\label{eq:nln}
\end{eqnarray}
Here $g(z)$ is given by Eq.~(\ref{g_final}), $dV/dz$ is the
cosmology--dependent comoving volume 
element per unity redshift interval and solid angle, and finally,
the cosmological parameter set $\tilde{q}_\alpha$ is given by
\begin{equation}
{\bf \tilde{q}}=\left\{\Omega_m,\Omega_{K},\Omega_b, h, M_\nu, n_s, w_0, w_a,
  \log(10^{10} A_s), A, B, \sigma_{\ln M,0}, \beta \right\}.
\label{qtilde_set}
\end{equation}
Where not otherwise specified, we marginalise over the four nuisance
parameters $A$, $B$, $\sigma_{\ln M,0}$, and $\beta$ to obtain the
constraints on $q_\alpha$ of Eq.~(\ref{q_set}) from cluster counts.
It is worth noting that in Eq.~(\ref{eq:fm_nc}) we have adopted the same approach of
Ref.~\cite{Shimon_etal2010}, i.e. we integrate the cluster distribution
over mass and consider number counts in redshift-space only, in order
to reduce the effects due to the model-dependence of the cluster mass inference.

Having defined the Fisher matrices $F_{\alpha\beta}$ for CMB,
$P_c(k)$, and $N_{i}$, respectively, the
1--$\sigma$ error on $q_\alpha$, marginalised over the
other parameters, is $\sigma(q_\alpha)=\sqrt{({F}^{-1})_{\alpha\alpha}}$.
Furthermore, to quantify the level of degeneracy between
the different parameters, we estimate the so-called
correlation coefficients, given by
\begin{equation}
r\equiv \frac{({F}^{-1})_{\alpha\beta}}
{\sqrt{({F}^{-1})_{\alpha\alpha}({F}^{-1})_{\beta\beta}}}.
\label{correlation}
\end{equation}
When the coefficient $|r|=1$ the two parameters are totally degenerate, while
$r=0$ means they are uncorrelated.

In \S~\ref{Results} we shall evaluate $\sigma(q_\alpha)$ and $r$ both
from galaxy cluster data, $F^{\rm P_c}_{\alpha\beta}$, $F^{\rm
  N}_{\alpha\beta}$, $F^{\rm P_c}_{\alpha\beta}+F^{\rm
  N}_{\alpha\beta}$,  
and from their combination with CMB priors, $F^{\rm
  P_c}_{\alpha\beta}+F^{\rm CMB}_{\alpha\beta}$, $F^{\rm
  N}_{\alpha\beta}+F^{\rm CMB}_{\alpha\beta}$,
$F^{\rm P_c}_{\alpha\beta}+F^{\rm N}_{\alpha\beta}+F^{\rm CMB}_{\alpha\beta}$.

\begin{table*} 
\caption{Forecast 1--$\sigma$ errors for the cosmological
  parameters considered in the text and the corresponding
  correlations $r$ with $M_\nu$, for
a \emph{Euclid}-like experiment alone and in combination with \emph{Planck}.}
\centering
\small
\setlength{\tabcolsep}{1.6pt}
\begin{tabular}{|l|c|c|c|c|c|c|c|c|c|c|c|c|}
\hline
\hline
\multicolumn{13}{c}{General cosmology}\\
\hline
{}&\multicolumn{2}{c}{\tiny{BAO}}&\multicolumn{2}{c}{\tiny{BAO+CMB}}&\multicolumn{2}{c}{\tiny{COUNTS${}^a$}}&\multicolumn{2}{c}{\tiny{COUNTS${}^a$+CMB}}&\multicolumn{2}{c}{\tiny{BAO+COUNTS${}^a$}}&\multicolumn{2}{c}{\tiny{BAO+COUNTS${}^a$+CMB}}\\
\hline
{}& $\sigma$&$r$ & $\sigma$&$r$ & $\sigma$&$r$ & $\sigma$&$r$ & $\sigma$&$r$ & $\sigma$&$r$\\
\hline
$ \Omega_{m}$&0.4139&0.7470& 0.0437&0.1278& 0.1560&0.8588& 0.0275&-0.3193& 0.0356&-0.0869& 0.0106&0.4046\\
$\Omega_{K} $&1.3327&-0.3479& 0.0302&0.1003& 2.6134&-0.2491& 0.0069&-0.2102& 0.4189&0.0849& 0.0039&0.3854\\
$ \Omega_{b}$&0.1008&0.6507& 0.0072&0.0083& 1.0042&-0.5785& 0.0050&-0.4842& 0.0143&-0.6459& 0.0016&0.0125\\
$h$         &0.3955&0.6583& 0.0562&-0.0158& 7.3604&0.5735& 0.0386&0.4771& 0.0411&-0.5120& 0.0125&-0.0380\\
$M_\nu$      &2.5972&1.0000& 0.2283&1.0000& 28.513&1.0000& 0.2564&1.0000& 0.8905&1.0000& 0.1853& 1.0000\\
$n_s$        &0.5332&-0.2732& 0.0024&-0.0666& 4.0365&-0.6573& 0.0024&-0.0486& 0.1452&0.8252& 0.0024&-0.0297\\
$w_0$        &1.7820&0.05685& 1.4648&-0.0248& 0.4659&0.0997& 0.2381&-0.2154& 0.3157&-0.2136& 0.2133&0.0020\\
$w_a$        &6.6788&-0.2651& 5.4101&-0.0044& 5.3758&-0.4089& 0.8187&0.0374& 1.0644&-0.2882& 0.7218&-0.2123\\
\footnotesize{$\log (10^{10}  A_{s})$}
             &--    &--     & 0.0253& 0.9117& 2.2631& 0.9057& 0.0287& 0.9336& 0.2899&0.1826& 0.0222&0.8878\\
\hline
\hline
\multicolumn{13}{c}{$\Lambda$CDM}\\
\hline
{}&\multicolumn{2}{c}{\tiny{BAO}}&\multicolumn{2}{c}{\tiny{BAO+CMB}}&\multicolumn{2}{c}{\tiny{COUNTS${}^a$}}&\multicolumn{2}{c}{\tiny{COUNTS${}^a$+CMB}}&\multicolumn{2}{c}{\tiny{BAO+COUNTS${}^a$}}&\multicolumn{2}{c}{\tiny{BAO+COUNTS${}^a$+CMB}}\\
\hline
{}& $\sigma$&$r$ & $\sigma$&$r$ & $\sigma$&$r$ & $\sigma$&$r$ & $\sigma$&$r$ & $\sigma$&$r$\\
\hline
$ \Omega_{m}$ & 0.1131&0.2559 & 0.0038&-0.8535 & 0.0461&0.7086 &0.0087&-0.9673 &
0.0252&0.3455 & 0.0036&-0.8361\\
$ \Omega_{b}$ & 0.0308&-0.2015 & 0.0010&-0.7291 & 0.6646&0.0628 &0.0021&-0.9401 &
0.0107&-0.4618 & 0.0009&-0.7042\\
$h$          & 0.0935&-0.2121 & 0.0071&0.8008 & 4.7404&0.3998 &0.0162&0.9561 &
0.0310&-0.4705 & 0.0068&0.7796\\
$M_\nu$      & 1.7791&1.0000 & 0.0818&1.0000 & 4.2971&1.0000 &0.1679&1.0000 &
0.5686&1.0000 & 0.0758&1.0000\\
$n_s$       & 0.3100&0.7894 & 0.0023&0.2498 & 1.4125&-0.6499 &0.0023&-0.0576 &
0.0713&0.5757 & 0.0023&0.2220\\
\footnotesize{$\log (10^{10}  A_{s})$}
            & --    & --    & 0.0140&0.6771 & 0.9446& 0.2623 &0.0225&0.9122 &
0.1021&0.8754 & 0.0138&0.7522\\
\hline
\hline
\end{tabular}
\begin{flushleft}
${}^a$\footnotesize{The four nuisance parameters, $A$, $B$,
    $\sigma_{\ln M,0}$, and $\beta$ have been marginalised over.}
\end{flushleft}
\label{err_corr}
\end{table*}
 \begin{table*}
\caption{Comparison of the \emph{Euclid}+\emph{Planck} forecast 1--$\sigma$ errors
  and the corresponding
  correlations with $M_\nu$, for different treatments of the nuisance parameters.}
\centering
\small
\setlength{\tabcolsep}{3.0pt}
\begin{tabular}{|l|cc|cc|}
\hline
\hline
\multicolumn{5}{c}{General cosmology}\\
\hline
{}&\multicolumn{2}{c}{\tiny{BAO+COUNTS${}^a$+CMB}}&\multicolumn{2}{c}{\tiny{BAO+COUNTS${}^b$+CMB}}\\
\hline
{} & $\sigma$ & $r$& $\sigma$ & $r$ \\
\hline
$ \Omega_{m}$&0.0106 & 0.4046 &  0.0057 & 0.6860 \\
$\Omega_{K} $&0.0039 & 0.3854 & 0.0031 & 0.6551 \\
$ \Omega_{b}$&0.0016 & 0.0125 & 0.0008 & 0.3456 \\
$h$         &0.0125 & -0.0380 & 0.0061 &-0.3841 \\
$M_\nu$      &0.1853 & 1.0000  & 0.1064 & 1.0000 \\
$n_s$       &0.0024 & -0.0297 & 0.0023 & 0.0722 \\
$w_0$       &0.2133 & 0.0020 & 0.1477 & 0.7178 \\
$w_a$       &0.7218 & -0.2123 & 0.5119 & -0.7616 \\
\footnotesize{$\log (10^{10}  A_{s})$}
            &0.0222 & 0.8878 & 0.0175 & 0.8599 \\
\hline
\hline
\multicolumn{5}{c}{$\Lambda$CDM}\\
\hline
{}&\multicolumn{2}{c}{\tiny{BAO+COUNTS${}^a$+CMB}}&\multicolumn{2}{c}{\tiny{BAO+COUNTS${}^b$+CMB}}\\
\hline
{} & $\sigma$ & $r$& $\sigma$ & $r$ \\
\hline
$ \Omega_{m}$ & 0.0036&-0.8361& 0.0008&0.1401\\
$ \Omega_{b}$ & 0.0009&-0.7042& 0.0003&0.5604\\
$h$          & 0.0068&0.7796& 0.0021&-0.3582\\
$M_\nu$       & 0.0758&1.0000& 0.0339&1.0000\\
$n_s$         & 0.0023&0.2220& 0.0023&0.5414\\
\footnotesize{$\log (10^{10}  A_{s})$}
              & 0.0138& 0.7522&0.0103&0.9441\\
\hline
\hline
\end{tabular}
\begin{flushleft}
${}^a$\footnotesize{The four nuisance parameters, $B_{M,0}$,
    $\sigma_{\ln  M,0}$, $\alpha$, and $\beta$ have been marginalised
    over.}\\
${}^b$\footnotesize{The four nuisance parameters, $B_{M,0}$,
    $\sigma_{\ln  M,0}$, $\alpha$, and $\beta$ have been kept fixed to
    the fiducial values reported in the text.}\\
\end{flushleft}
\label{nuis_comp}
\end{table*}
\begin{figure*}
\centering
\begin{tabular}{l l}
\includegraphics[width=7.3cm]{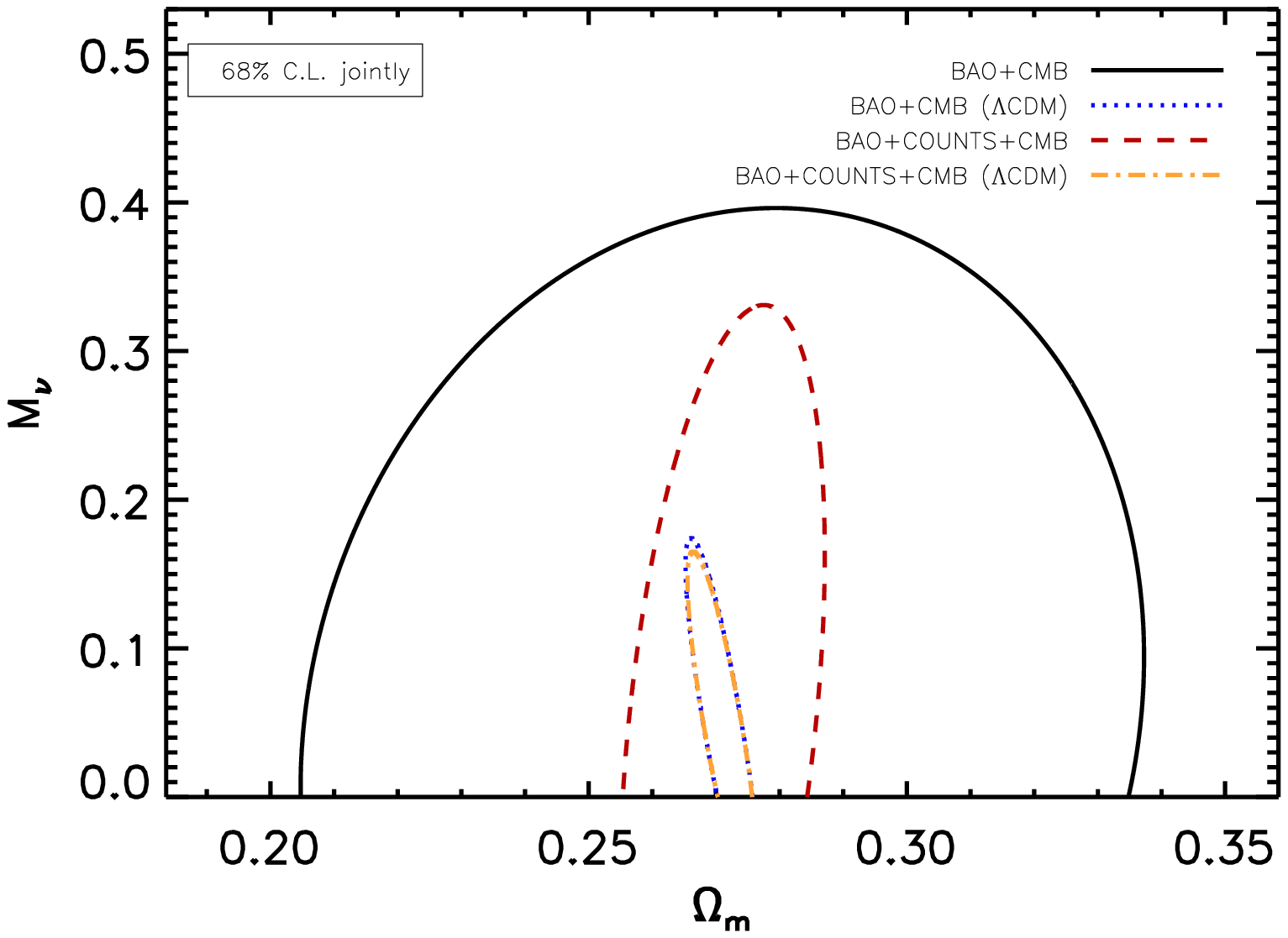}&
\includegraphics[width=7.3cm]{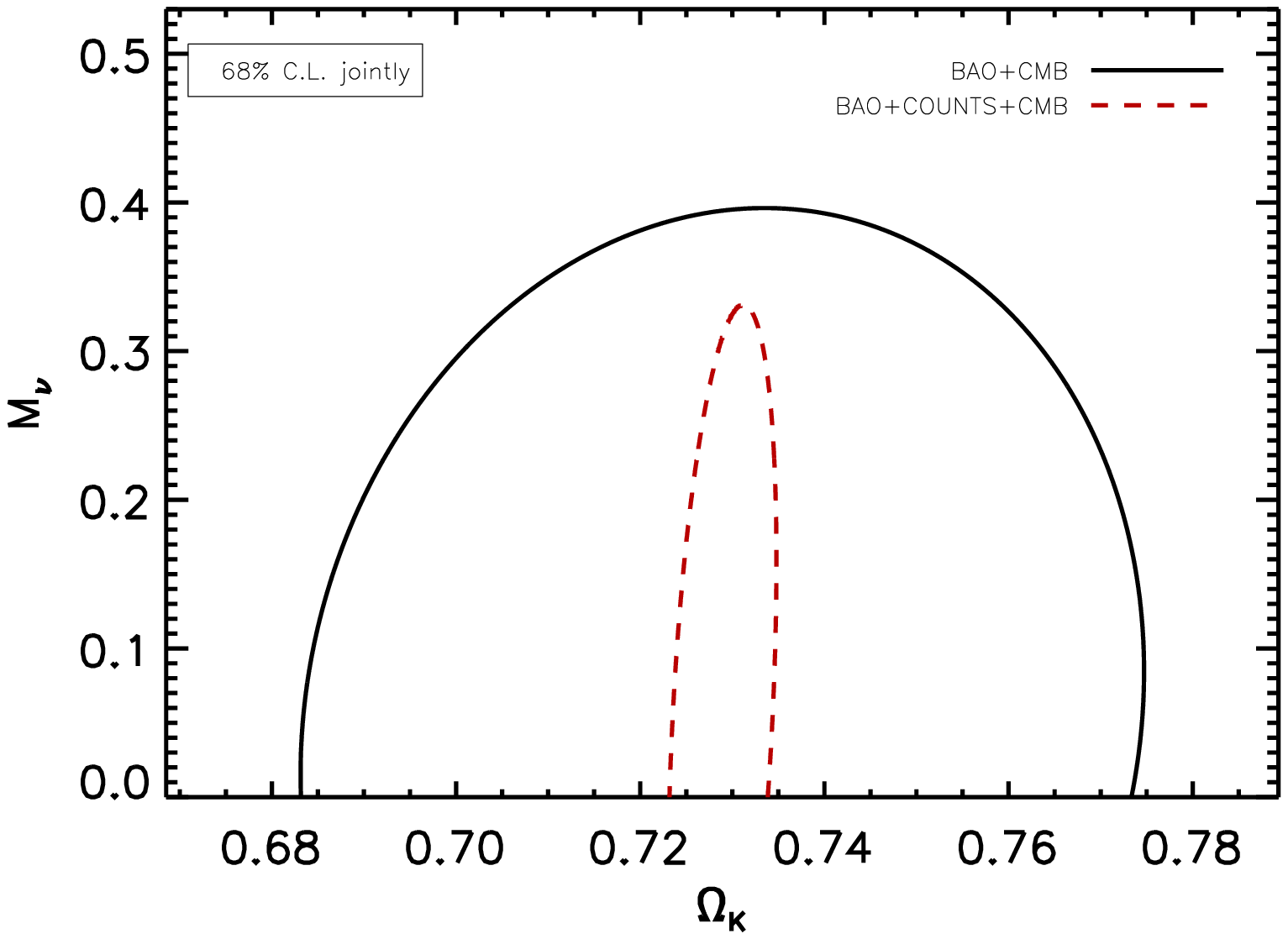} \\
\includegraphics[width=7.3cm]{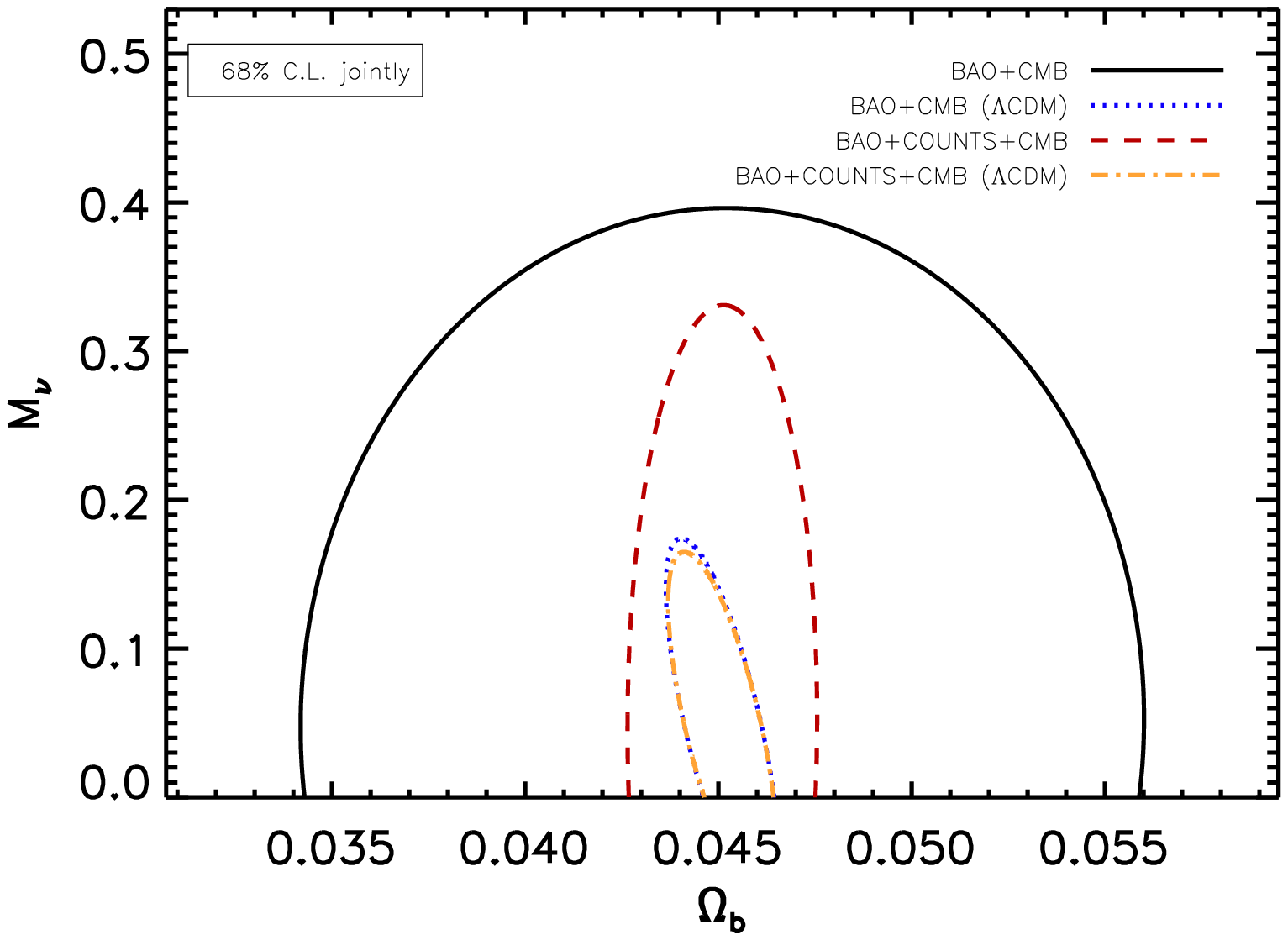}&
\includegraphics[width=7.3cm]{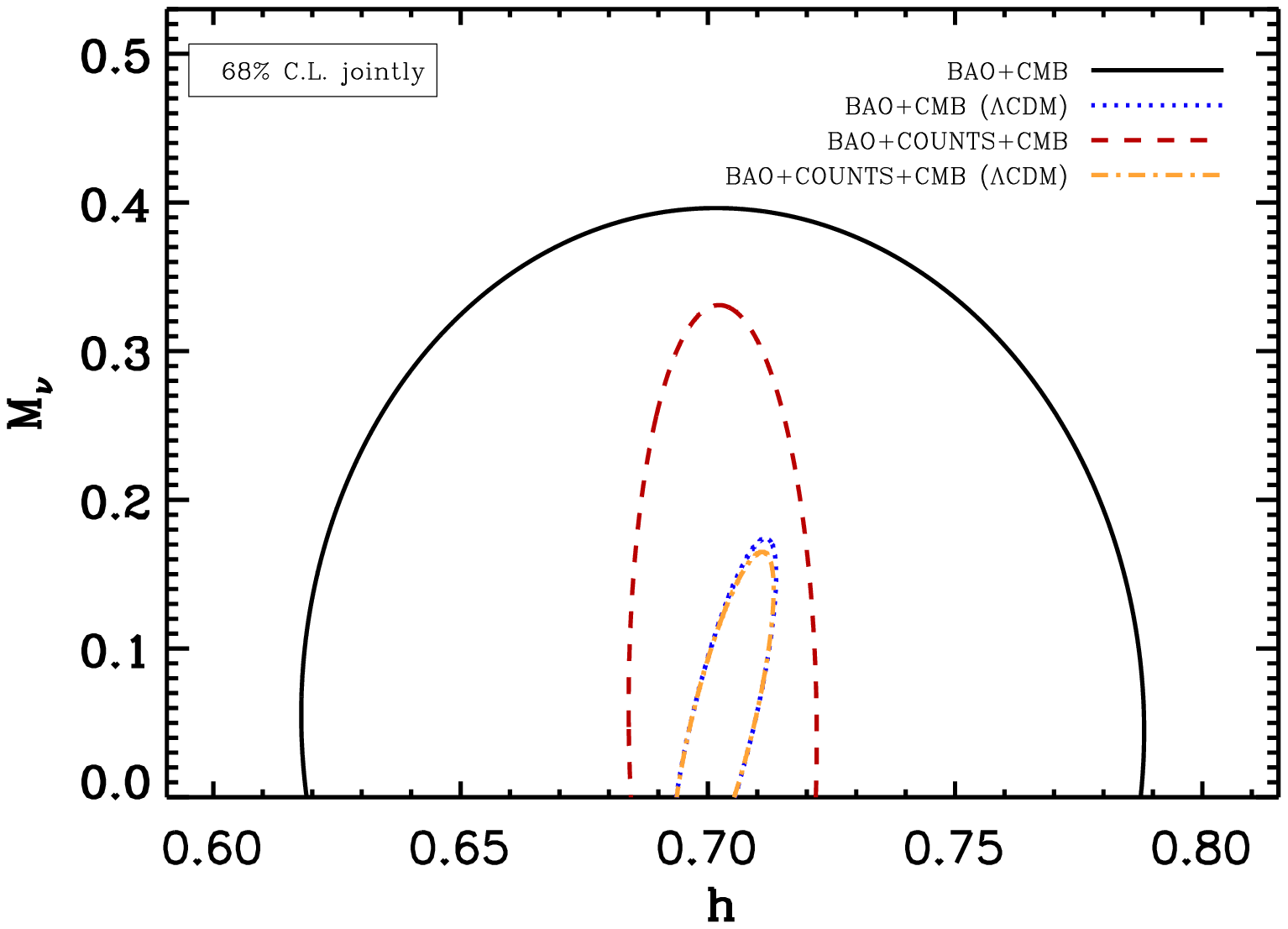}\\
\end{tabular}
\caption{2-parameter projected $68\%$ C.L. contours in the
$M_\nu$-$q_\alpha$ subspace with $q_\alpha=\Omega_m,\,\Omega_K,\,\Omega_b,\,h$. The solid black line and the dotted blue
line correspond to BAO+$P_{\rm c}(k)$-shape
cluster data from a \emph{Euclid}-like survey in combination with \emph{Planck}
priors for a general cosmology, and a $\Lambda$CDM Universe,
respectively. The dashed red line, and the dot-dashed orange line are
obtained with the further addition of cluster counts data, and
represent the confidence contours obtained for the two cosmological
models, respectively.}
\label{contour1}
\end{figure*}
\begin{figure*}
\centering
\begin{tabular}{l l}
\includegraphics[width=7.3cm]{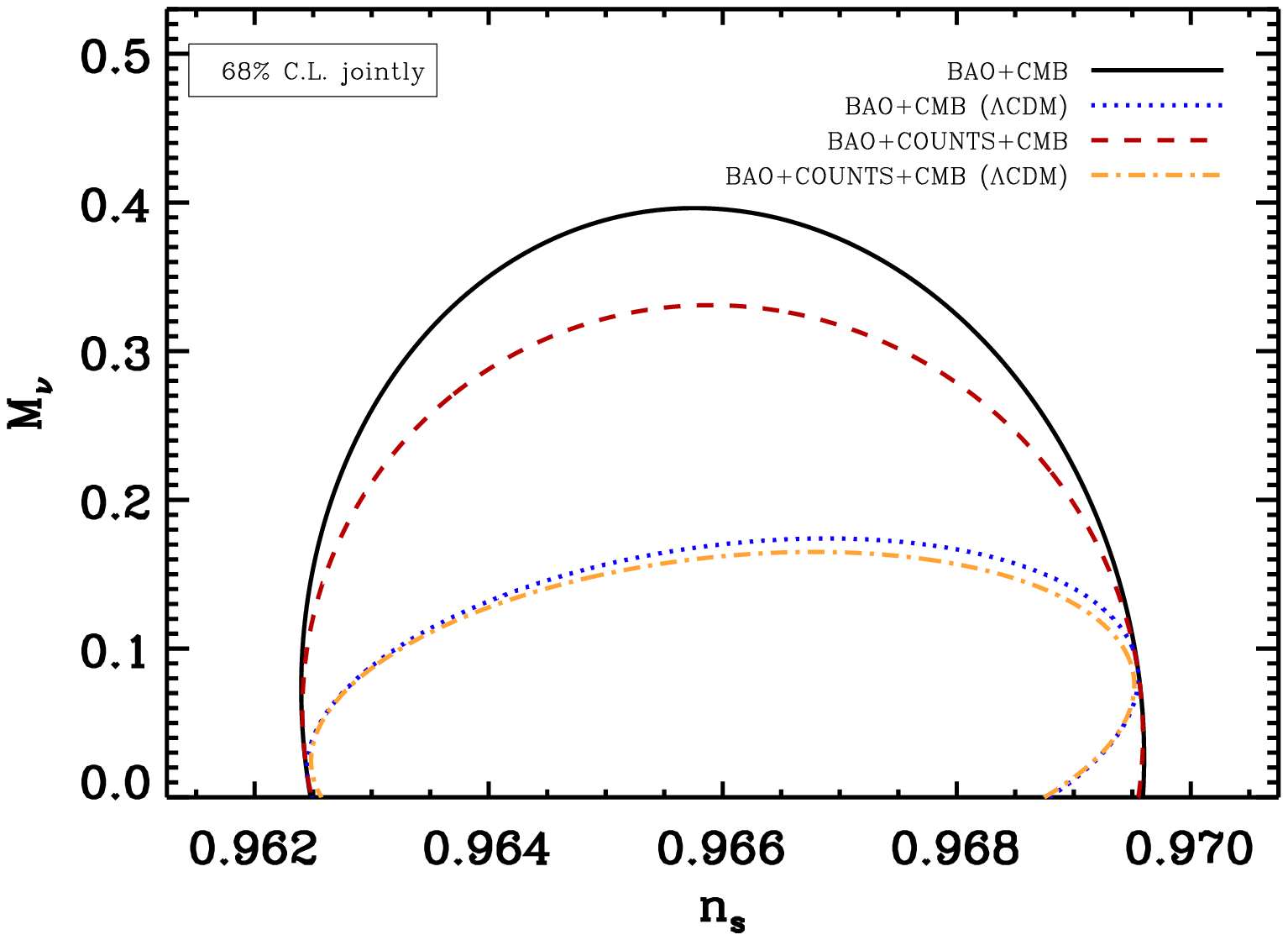}&
\includegraphics[width=7.3cm]{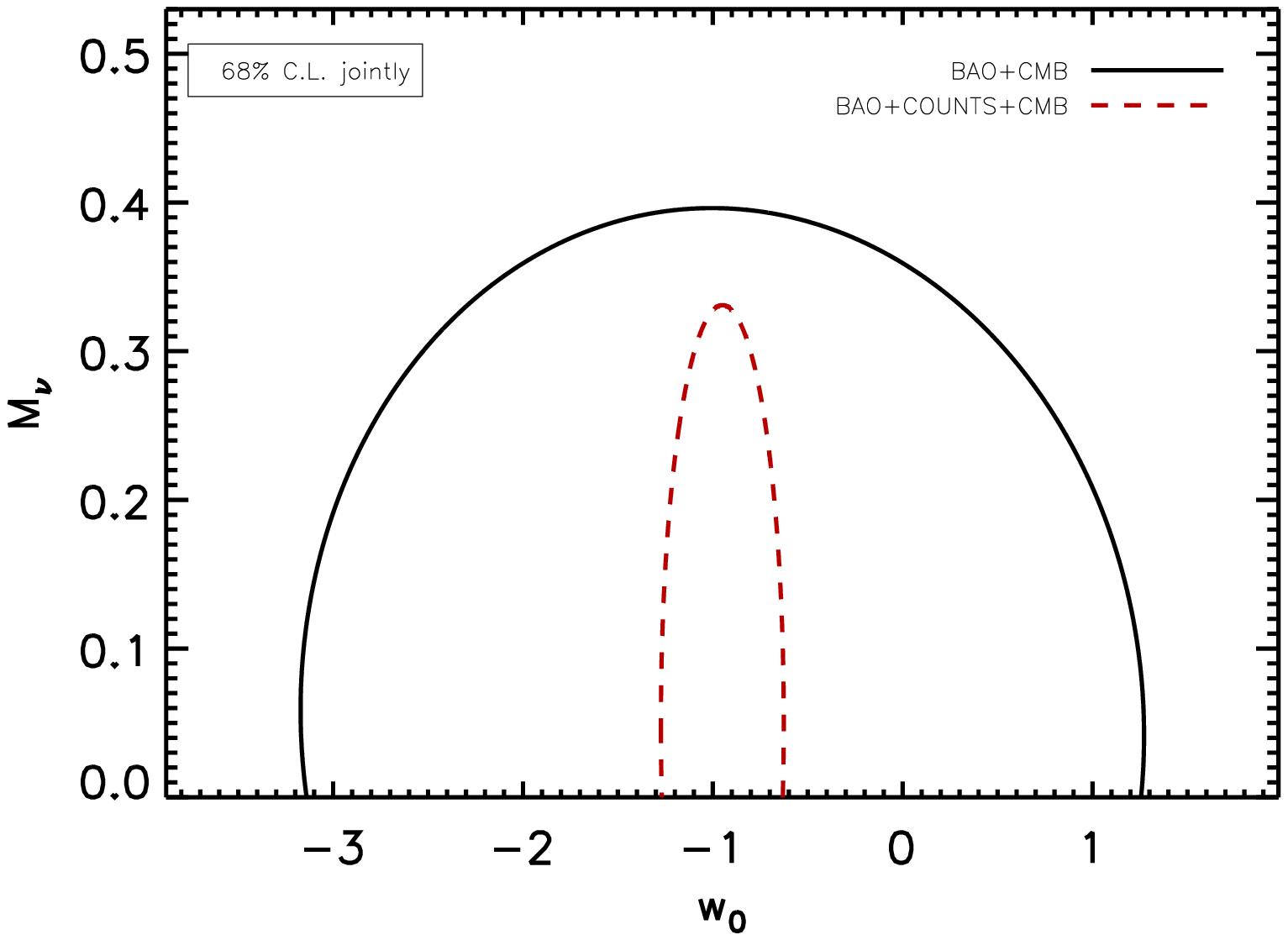} \\
\includegraphics[width=7.3cm]{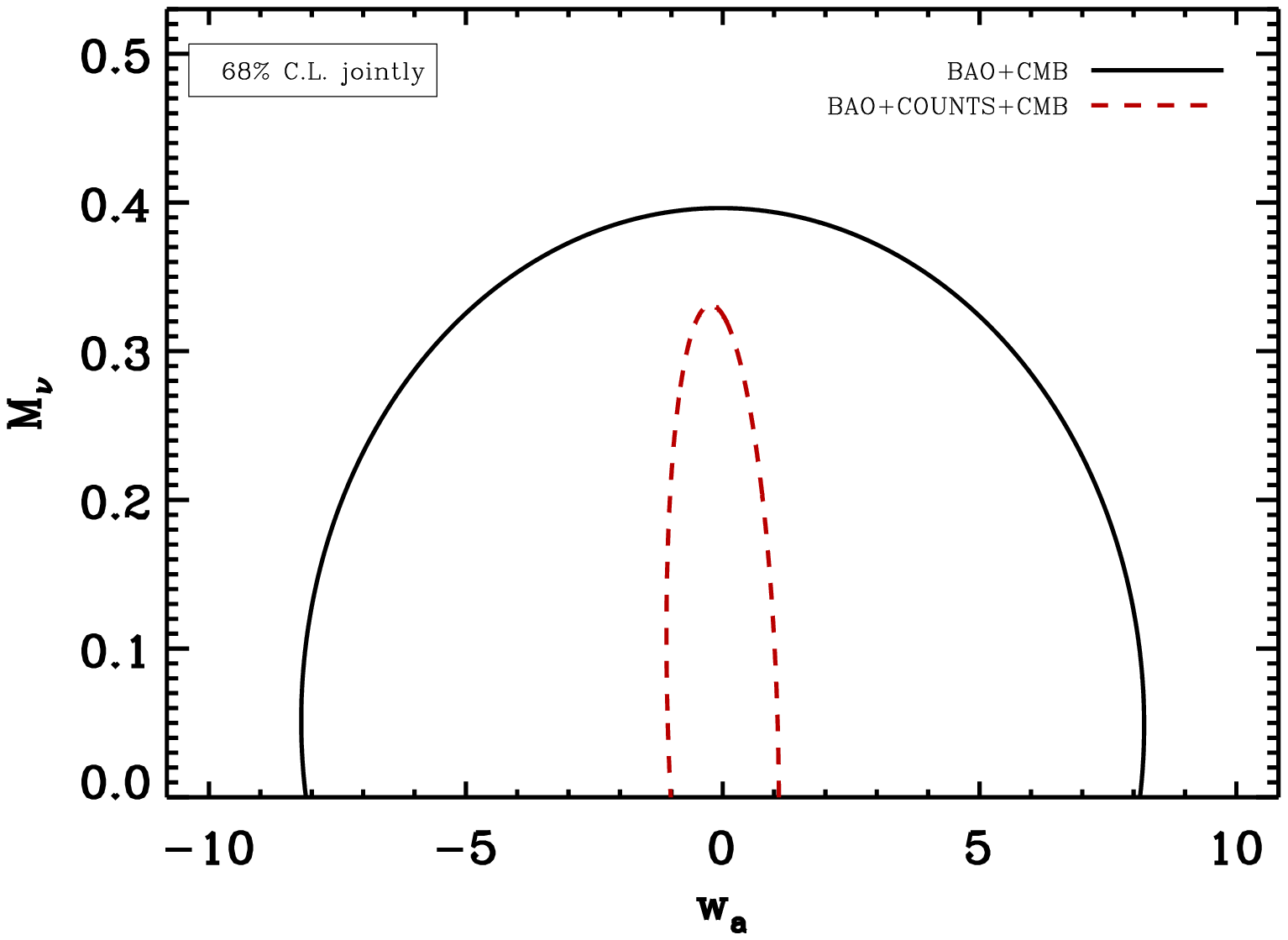}&
\includegraphics[width=7.3cm]{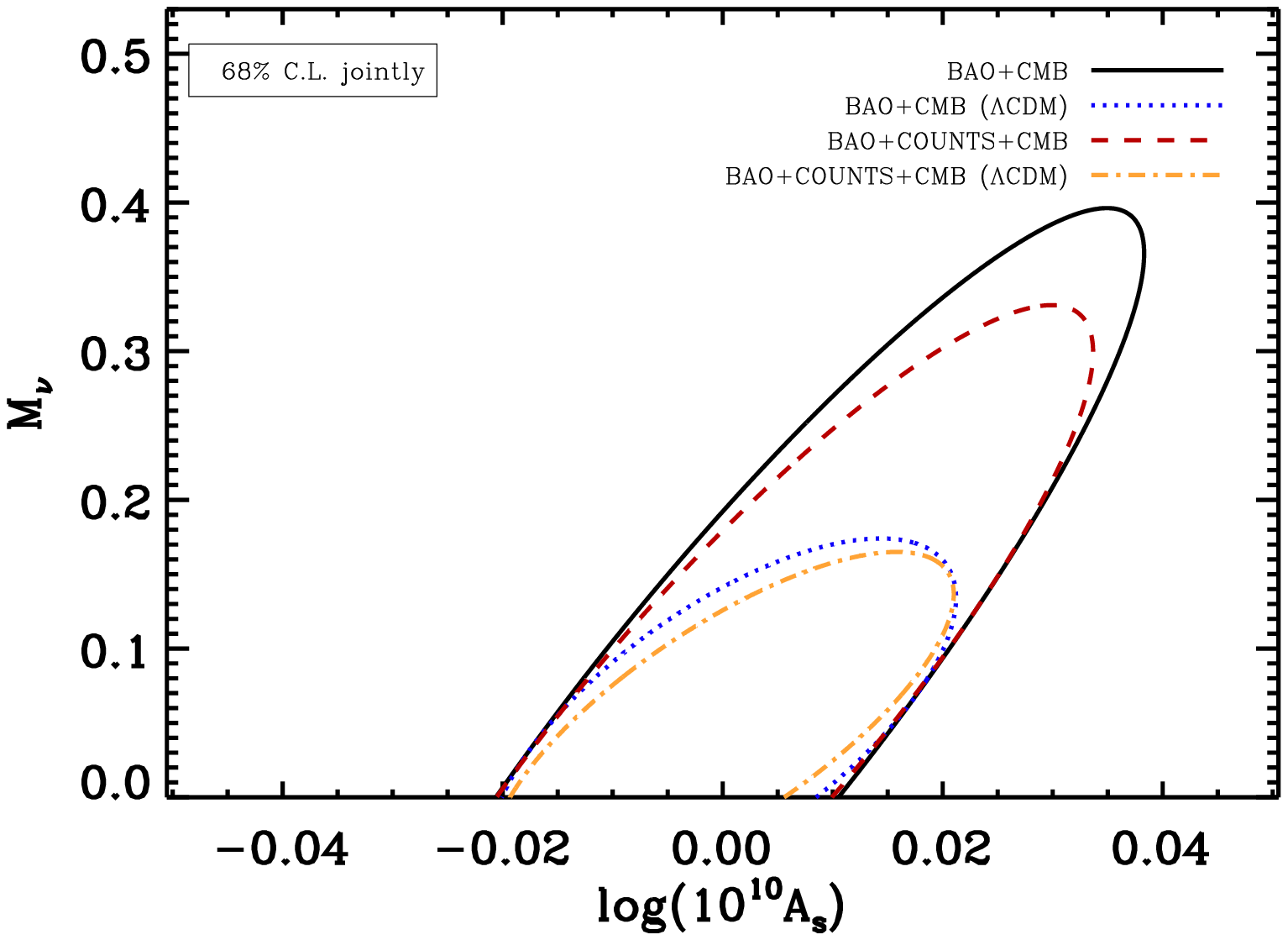}\\
\end{tabular}
\caption{2-parameter projected $68\%$ C.L. contours in the $M_\nu$-$q_\alpha$ subspace with
$q_\alpha=n_s,\,w_0,\,w_a,\,\log(10^{10}A_s)$. The solid black line and the dotted blue      
line correspond to BAO+$P_{\rm c}(k)$-shape cluster data from a
\emph{Euclid}-like survey in combination with \emph{Planck} priors for a general
cosmology, and a $\Lambda$CDM Universe, respectively. The dashed red
line, and the dot-dashed orange line are obtained with the further
addition of cluster counts data, and represent the confidence contours
obtained for the two cosmological models, respectively.}
\label{contour2}
\end{figure*}

\subsection{\emph{Euclid}-like survey}
\label{Survey}
In this work we forecast neutrino constraints using count and power
spectrum measurements of galaxy clusters photometrically selected by a
future survey like the one planned by the ESA selected M-class \emph{Euclid} galaxy survey.
This survey will be able to push towards very high redshifts over a large area,
thanks to its unique capabilities in the infrared which cannot be
matched from the ground. Conservative 
estimates, based on simulated mock catalogues, indicate that \emph{Euclid}
will find of order $\sim 6 \times 10^4$ clusters with S/N better than
$3$, between $z=0.2$ and $z=2.0$, with $10^4$ at $z>1$ \cite{RedBook}. In this case, the cluster-based constraints on cosmological parameters
will be limited by the understanding of the catalogue selection
function, systematic errors and cluster mass determinations and their
uncertainties. In this respect, \emph{Euclid} will be able to
calibrate the mass-observable relations and their scatter through
lensing measurements. The high image quality and number density of sources will
enable \emph{Euclid} to measure masses of 
clusters much more accurately and out to higher redshifts than is
possible from the ground. Moreover, the combination of \emph{Euclid}
data with other surveys, such as
eROSITA\footnote{http://www.mpe.mpg.de/erosita/} or
Planck, will enable the cross-calibration of non-lensing mass-observable
relations, which are currently limited to low redshifts and small 
samples. For instance, \emph{Euclid} will provide mass proxies via the
stacking of clusters according to their X-ray or Sunyaev-Zel'dovich signals.
\cite{RedBook}.
Here we adopt the following specification for the \emph{Euclid}-like galaxy
cluster catalogue analysed in this work:
area$=15 \times 10^3 {\rm deg^2}$, $z_{\rm min}=0.2$, $z_{\rm max}=2$,
$\Delta_z=0.1$, S/N$>5$.

\section{Results}
\label{Results}
In this Section we present the predicted 1--$\sigma$ marginalised
errors and correlations for the cosmological parameter set, described
in Eq.~(\ref{q_set}), considered in
this work, focusing in particular on the total neutrino mass $M_\nu$. As explained
in \S~\ref{Fiducial cosmologies}, we have assumed an inverted mass
hierarchy with a fiducial $M_\nu=0.05$ eV, which represents the minimum total neutrino
mass predicted by neutrino oscillation experiments. In this respect, we
have been very conservative, and consequently the forecast
$\sigma(M_\nu)$ errors should be viewed as upper bounds for neutrino
mass measurements in the context of a \emph{Euclid}-like galaxy cluster
survey in combination with \emph{Planck}.

In Table~\ref{err_corr} we show the marginalised errors for the
cosmological parameters of Eq.~(\ref{q_set}) and the corresponding
correlation coefficients with $M_\nu$, for the different probes and
their combinations analysed in this work. Moreover,
Figs.~\ref{contour1}-\ref{contour2} represent the corresponding 2-parameter $M_\nu$-$q_\alpha$ joint
contours at $68\%$ C.L. Finally, in Table~\ref{nuis_comp} we compare the cosmological
parameter errors and their correlations with $M_\nu$ as obtained from
galaxy cluster measurements in combination with CMB data, for the two different cases in which the
nuisance parameters $B_{M,0}$, $\sigma_{\ln  M,0}$, $\alpha$ and
$\beta$ are kept fixed to their fiducial values or are marginalised over.

\subsection{BAO+CMB}
\label{bao_cmb}
Let us consider the different probes separately. 
The $2^{\rm nd}$ and $3^{\rm rd}$ columns of the upper panel of Table~\ref{err_corr} show
the 1--$\sigma$ errors and the values of the $r$ coefficients for the $M_\nu$--$q_\alpha$
correlations obtained in a general cosmology which includes
dark energy and curvature, and exploiting information from the combination
of BAO+$P(k)$-shape extracted from the observed galaxy cluster power
spectrum of Eq.~(\ref{eq:Pobs_here}).
As stated in Appendix \ref{appendix:Pobs}, we assume that up to $z=1$
photometrically selected clusters are also spectroscopically confirmed, and
consequently we assume a redshift error $\sigma_z=0.001 (1+z)$ for
$0.2<z<1$, and $\sigma_z=0.03 (1+z)$ for $1<z<2$. We have verified
that in this case $99.95 \%$ of the 
signal comes from $z<1$, for both neutrino mass and dark energy
parameters $w_0$ and $w_a$. This is due to the damping effect on the
observed power spectrum caused by the redshift error $\sigma_z$
  which increases linearly with redshift (see
  Eq.~(\ref{eq:Pm})). Supposing instead that 
photometrically 
selected galaxy clusters could be spectroscopically confirmed up
to $z=2$, this would imply $96 \%$ of the signal coming from $z<1$ 
as regards neutrino mass constraints, while for errors on the
dark energy equation of state, one would get the $99 \%$ of the signal
in terms figure of merit from $z<1$. After 2019, future galaxy surveys
will probably be able to reach spectroscopic precision on $\sigma_z$ even for
redshifts $z>1$. 

As we can observe, the marginalised errors of all
the cosmological parameters are $1$--$2$ orders of magnitude larger
than the corresponding errors obtained from galaxy power spectrum measurements
\cite{Melita11}. This is expected and due to the fact that the galaxy
cluster spatial density
is much lower than the density of spectroscopically selected galaxies from
a \emph{Euclid}-like survey, and evidently this effect is not compensated by the larger
value of the effective bias. Let us notice that, having marginalised
over the cluster power spectrum amplitude and redshift-space
distortions, the information from BAO+$P_{\rm c}(k)$-shape 
does not provide any constraint on $\log(10^{10}A_s)$.
In this case the 1--$\sigma$ error on the total neutrino mass is quite large,
$\sigma(M_\nu)=2.6$ eV and, except for $w_0$, all the cosmological
parameters are non-negligibly correlated with $M_\nu$, with the main
correlations given by $M_\nu-\Omega_m$, $M_\nu-\Omega_b$,
and $M_\nu-h$. 

Looking at the $4^{\rm th}$ and $5^{\rm th}$ columns of the upper
panel of Table~\ref{err_corr}, we can observe that all the constraints improve considerably, by $1-2$ orders
of magnitude, when \emph{Planck} priors
are added to BAO+$P_{\rm c}(k)$-shape cluster data, except for the
dark energy equation of state, owing to the CMB weakness in
constraining $w_0$ and $w_a$ simultaneously. Focusing on the
neutrino mass, in this case $\sigma(M_\nu)=0.23$ eV and all the
parameter degeneracies with $M_\nu$  are broken, except for
$\log(10^{10}A_s)$, which is now constrained by CMB measurements and
is highly degenerate with $M_\nu$, $r=0.9$, since these two parameters
produce opposite effects on the cluster power spectrum.

Let us now consider a less general cosmology, i.e. the
$\Lambda$CDM model, where the curvature $\Omega_K$ and the dark energy
equation of state are kept fixed to their fiducial values. Looking at
the $2^{\rm nd}$ column of the lower panel in Table~\ref{err_corr}, we note
that the 1--$\sigma$ error on the neutrino mass decreases by a factor
of $\sim 1.5$ with respect to the general case, and the combination with \emph{Planck}
priors ($4^{\rm th}$ column) gives $\sigma(M_\nu)=0.08$ eV, which is comparable with the
lowest value of the total neutrino mass, $M_\nu=0.05$ eV, predicted by
neutrino oscillation experiments.  This means that the combination of
CMB data with BAO+$P_{\rm c}(k)$-shape measurements from galaxy
clusters, selected from future nearly all-sky galaxy catalogs, could
detect values of the neutrino mass close to the minimum one assuming a
$\Lambda$CDM Universe. Anyway, all the cosmological parameters keep a
large correlation 
with $M_\nu$, implying that information from external data, as e.g. a
Gaussian prior on $h$, could help to further improve the neutrino mass
constraint. 

\subsection{COUNTS+CMB}
\label{counts_cmb}
The cosmological parameter forecasts obtained from cluster counts in a
\emph{Euclid}-like galaxy survey are reported in the $6^{\rm th}$ and $7^{\rm th}$
columns of the upper panel in Table~\ref{err_corr} for a general
cosmology, and in the corresponding columns of the lower panel for the
$\Lambda$CDM case. The four nuisance parameters described in
\S~\ref{Nuisance} have been marginalised over. The constraining power
of cluster counts depends on the considered parameter: for example the
1--$\sigma$ errors on $\Omega_m$ and $w_0$ are respectively $\sim 2.5$
and $\sim 4$ times lower than the ones obtained from
cluster BAO+$P_{\rm c}(k)$-shape measurements, for both the model
cosmologies. Nonetheless, the constraints 
on all the remaining cosmological parameters are much worse than the
BAO+$P_{\rm c}(k)$-shape forecasts, and in particular the error on the
total neutrino mass from cluster counts for the general cosmology,
$\sigma(M_\nu)=28.5$ eV, is $\sim 11$ times larger than from the cluster
power spectrum, while, for the $\Lambda$CDM model it is $\sim 2.5$ times larger,
$\sigma(M_\nu)=4.3$ eV. For cluster counts as well, all the
cosmological parameters are highly correlated with $M_\nu$ and the
larger correlation coefficients $r$ are given by $M_\nu-\Omega_m$,
$M_\nu-\Omega_b$, $M_\nu-h$, $M_\nu-n_s$, and
$M_\nu-\log(10^{10}A_s)$.

Adding cluster counts data to \emph{Planck} priors breaks
degeneracies in CMB measurements, especially for $\Omega_K$, $w_0$ and
$w_a$, so that cosmological parameter constraints coming from
their combination improve considerably when compared to cluster counts
alone. Anyway, also in this case, all the cosmological parameters remain
strongly correlated with the neutrino mass. For COUNTS+CMB we find
$\sigma(M_\nu)=0.26$ eV for a cosmological model which includes
curvature and an evolving dark energy component, and
$\sigma(M_\nu)=0.17$ eV for a 
$\Lambda$CDM cosmology. These findings have to be compared with the
corresponding errors from cluster BAO+$P_{\rm c}(k)$-shape
measurements, given by $\sigma(M_\nu)=0.23$ eV and
$\sigma(M_\nu)=0.08$ eV, respectively.

\subsection{BAO+COUNTS}
\label{bao_counts}
Being complementary probes, the separate
constraining powers of cluster counts and cluster BAO+$P_{\rm
  c}(k)$-shape data are notably amplified when combined
together. In the case of the present work, the degeneracy breaking,
provided by such a combination, is mainly due to the inclusion, in
cluster count measurements, of information from the power spectrum amplitude
and from the growth of structure (encapsulated in $n(M,z)$), which, on
the contrary, are marginalised over for BAO+$P_{\rm c}(k)$-shape
measurements. This additional information helps 
to mitigate the degeneracies between the cosmological
parameters, present when each probe is taken individually. 
Nonetheless, it is worth noting that $dN/dz$ and $P_c(k,z)$ describe different
phenomena, abundance and clustering respectively, and have a different dependence on
cosmological parameters; for instance, the growth rate of perturbations $f_g$
influences mainly the redshift evolution of cluster abundance, while
free-streaming of neutrinos suppresses fluctuation power on small
scales, mainly affecting the cluster correlation. Therefore, the inclusion
even of amplitude and growth information in cluster power spectrum
measurements would improve the cosmological parameter errors, but, anyway,
would not replace the role of cluster counts in statistical
constraints \cite{0406331}. Moreover, these two probes are affected by
totally different systematics, and this contributes in breaking the
degeneracies present in each method.

The qualitative explanation above can be verified by
comparing the $10^{\rm th}$ column of the upper panel
of Table~\ref{err_corr} with the $2^{\rm nd}$ and $6^{\rm th}$
columns. We notice how the 1--$\sigma$ errors for
$\Omega_m$, $\Omega_{K}$, $\Omega_b$, $h$, and $\log(10^{10} A_s)$ are
reduced by $\sim 1$ order of magnitude with
respect to the minimum of the corresponding errors from cluster
counts and cluster BAO+$P_{\rm c}(k)$-shape measurements taken separately. The
decrease is less pronounced but still important for $n_s$, $w_0$,
$w_a$. In particular for the total neutrino mass we find
$\sigma(M_\nu)=0.9$ eV in a general model cosmology, which is
comparable to constrains from the Ly$\alpha$ forest at $95 \%$
C.L. \cite{Matteo1}, and 
$\sigma(M_\nu)=0.57$ eV in a $\Lambda$CDM cosmology (lower panel of
Table~\ref{err_corr}), comparable to 
constraints from redshift space distortions \cite{Marulli_etal2011a}.
The cosmological parameters more strongly correlated with $M_\nu$ are
in this case $\Omega_b$, $h$, and $\log(10^{10} A_s)$.

\subsection{BAO+COUNTS+CMB}
\label{bao_counts_cmb}
In this Section we consider the total combination of cluster counts and
BAO+$P_{\rm c}(k)$-shape measurements with CMB data. The results are
shown in the last two columns of Table~\ref{err_corr}, while Figs.~\ref{contour1}-\ref{contour2}
represent the 2-parameter projected $68\%$ C.L. contours in the
$M_\nu$-$q_\alpha$ subspace. The solid black line and the dotted blue
line correspond to BAO+$P_{\rm c}(k)$-shape
cluster data from a \emph{Euclid}-like survey in combination with \emph{Planck}
priors for a general cosmology, and a $\Lambda$CDM Universe,
respectively. The dashed red line, and the dot-dashed orange line are
obtained from the further addition of cluster counts data, and
represent respectively the confidence contours obtained for the two cosmological
models.
These plots clearly display the contribution of cluster counts in
constraining cosmological parameters, and show that the main
effect results from breaking degeneracies when the cosmological model
includes curvature and an evolving dark energy component (compare the
solid black 
line with the red dashed one), so that the 1--$\sigma$ errors decrease
by $\sim 4-7$ times with respect to the combination CMB+BAO+$P_{\rm
  c}(k)$-shape for all the parameters, except for $n_s$, $\log(10^{10}
A_s)$, and $M_\nu$. Specifically, for the total neutrino mass error in
a general cosmology we find $\sigma(M_\nu)=0.18$ eV, which reduces to
$\sigma(M_\nu)=0.076$ eV in a $\Lambda$CDM Universe\footnote{For
  completeness and comparison reasons, we have considered also the
  case where only photometric redshifts are available. Under this
  assumption, for the combination CMB+BAO+$P_{\rm
  c}(k)$-shape we find $\sigma(M_\nu)=0.24$ eV in a general cosmology,
  which reduces to $\sigma(M_\nu)=0.16$ eV in a $\Lambda$CDM Universe,
  i.e. neutrino mass constraints would worsen by a factor $\sim 1.3$
  and $\sim 2$, respectively, if a spectroscopic follow-up of galaxy
  cluster up to $z=1$ would not be feasible.}. It is worth noting 
that, in the latter case, all the cosmological parameter constraints are
comparable to the corresponding ones as obtained in a general
cosmology from the galaxy 
CMB+BAO+$P_{\rm g}(k)$-shape measurements in a \emph{Euclid}-like spectroscopic
galaxy survey \cite{Melita11}.  This means that, even if the galaxy
clustering constraining power
is undoubtedly superior to the galaxy cluster one, the two probes are
complementary and their combination could greatly help to improve
constraints on all the cosmological parameters, including neutrino
mass and dark energy.

Finally, the $2^{\rm nd}$ and $4^{\rm th}$ columns of
Table~\ref{nuis_comp} show that, fixing the four
nuisance parameters $A$, $B$, $\sigma_{\ln M,0}$, and $\beta$, the
1--$\sigma$ errors on the total neutrino mass reduce from $0.18$ eV to
$0.11$ eV for a general cosmology, and from $0.076$ eV to $0.03$ eV for
a $\Lambda$CDM model. Therefore, if the true underlying
cluster masses were known without uncertainties, 
it would be possible to detect at 3--$\sigma$ level and measure the
mass of cosmic neutrinos in a cosmology-independent way, if the sum of neutrino
masses is above $0.3$ eV, while assuming spatial flatness and
cosmological constant otherwise. 

\section{Conclusions}
\label{Conclu}

In this work we adopted a Fisher matrix approach to forecast
constraints on a wide array of cosmological parameters from the
photometric galaxy cluster catalog produced by future
\emph{Euclid}-like surveys. Specifically,  we focused attention on the
total mass of neutrinos as derived by a combination of cluster number
counts and the shape of the cluster power spectrum, including BAO
information. In addition to the cosmological parameters of the
standard $\Lambda$CDM model and the total neutrino mass, we also
considered non-vanishing curvature, as well as a dynamical evolution
of the dark energy equation of state. We marginalised our constraints
over the growth function, the amplitude of the scalar curvature power
spectrum, and a set of nuisance parameters intended to represent the
uncertainties in assigning true masses to galaxy clusters. In order to
make our analysis more complete, we also added priors on cosmological
parameters coming from the ongoing CMB experiment \emph{Planck}. Our
main results can be summarised as follows. 
\begin{itemize}
\item Constraints coming from cluster counts and power spectrum
  shape+BAO separately are in general relatively weak. Errors on
  parameter estimated using the cluster correlation function alone are up to
  two orders of magnitude larger than what is obtained with the galaxy
  correlation function, due to the lower spatial number density of
  clusters. Number counts alone do better than this only for the
  matter density parameter and the present-day dark energy equation of
  state, while they perform worst in all other cases. Specifically,
  the error on the total neutrino mass is $\sigma(M_\nu) = 2.6$ eV for
  cluster correlation function alone and $\sigma(M_\nu) = 28.5$ eV for
  cluster counts alone in the context of a general cosmological model
  including curvature and evolving dark energy.
\item Combining the cluster power spectrum shape+BAO with cluster
  counts exploits the advantages of both approaches, and hence greatly
  improves the constraints of individual probes. The errors on
  cosmological parameters estimated by the combination of the two
  tests are reduced in general very substantially for all parameters,
  and by up to an order of magnitude for some of them. For a general
  cosmology, the error on
  the total neutrino mass is brought down to $\sigma(M_\nu) =
  0.9$ eV, comparable with present constraints from the
Ly$\alpha$ forest \cite{Matteo1}
  and CMB in non-flat models \cite{Smith_etal2011}. 
\item Constraints on cosmological parameters are also greatly
  tightened by either $i)$ adopting \emph{Planck} priors, and $ii)$
  reducing the number of free parameters by shrinking the cosmological
  model to a standard $\Lambda$CDM one. For a general cosmology, adding \emph{Planck} priors to
  the cluster power spectrum shape+BAO reduces the parameter errors by
  $1-2$ orders of magnitude, except for the dark energy parameters,
  while performing the same operation on the cluster counts also
  improves the constraining power by more than one order of
  magnitude. Moreover, these priors help to remove many of the
  degeneracies between cosmological parameters, thus reducing the
  error on the total neutrino mass from either probes down to
  $\sigma(M_\nu)\sim 0.25$ eV, which goes down to $\sigma(M_\nu) \sim
  0.08-0.17$ if only a $\Lambda$CDM background is considered.
\item The combination of cluster counts, power spectrum shape+BAO, and
  \emph{Planck} priors produces very competitive constraints. In a
  $\Lambda$CDM context, the
  errors on all cosmological parameters are compatible with those
  derived for a general cosmology from the galaxy power spectrum shape+BAO performed for
  the spectroscopic catalog of a future \emph{Euclid}-like
  survey in combination with \emph{Planck}. Errors on the total neutrino mass are now down to
  $\sigma(M_\nu) = 0.18$ eV, and $\sigma(M_\nu) = 0.076$ eV if the
  cosmological model is enforced to be standard $\Lambda$CDM. The
  latter value is only $1.5$ times the minimum neutrino mass admitted
  by neutrino oscillation experiments. 
\item Fixing the value of the nuisance parameters, which equals
  assuming a perfect knowledge of cluster true masses, also greatly
  improves parameter constraints. It is quite unlikely that this level
  of precision can be reached for all clusters in the \emph{Euclid}
  photometric catalog, while it might be a reasonable assumption for a
  subsample of objects having an extensive multi-wavelength
  follow-up. Nonetheless, if this will indeed be the case, the errors
  on all estimated parameters would be reduced by up to a factor of a
  few. Specifically, the error on the total neutrino mass would go
  down to $\sigma(M_\nu) = 0.1$ eV for the general cosmology and to only
  $\sigma(M_\nu) = 0.034$ eV for the $\Lambda$CDM one, which means
  that the minimum neutrino mass could be detected. 
\end{itemize}
\noindent
We emphasize that the analysis presented here is quite conservative from
the point of view of the cosmological information, in that it does not
include any guidance from the growth function in the observed galaxy cluster
power spectrum, the amplitude of the
scalar curvature power spectrum, and the redshift space
distortions. At the same time, we assumed throughout this work a
Gaussian distribution of initial density and curvature
perturbations. It has been repeatedly shown that the primordial
non-Gaussianity has a substantial impact on both the number counts of
massive galaxy systems and their correlation function 
\cite{Slosar_etal2008,Grossi1,Grossi2,Grossi3,Pillepich,Desjacques,Dalal,MaggioreRiotto,MV08,Sheth,Damico,McDonald}. Inclusion
of this primordial non-Gaussianity would have the effect of adding at
least another parameter to the study, hence loosening somewhat the
constraints that have been found here \cite{olga}. 

The present investigation shows in a neat way that, whilst taken
at face value the galaxy correlation function is superior to the
cluster correlation function, the combination of the two, and of the
latter with cluster number counts, can be of invaluable help in
breaking the degeneracies between, and hence effectively improving
the constraints on, cosmological parameters. Therefore galaxy
clusters are expected to play a key role for precision cosmology
with future large galaxy surveys, addressing even questions related
to fundamental physics such as the value of the neutrino mass. 

\acknowledgments
We acknowledge S.~Bardelli for useful discussions.
We acknowledge financial contributions from contracts ASI-INAF
I/023/05/0, ASI-INAF I/088/06/0, ASI I/016/07/0 'COFIS', ASI
'\emph{Euclid}-DUNE' I/064/08/0, ASI-Uni Bologna-Astronomy
Dept. '\emph{Euclid}-NIS' I/039/10/0, and PRIN MIUR 'Dark energy
and cosmology with large galaxy surveys'.

\appendix

\section{The observed galaxy cluster power spectrum}
\label{appendix:Pobs}
The Fisher matrix is defined as the
second derivative of the natural logarithm of the likelihood surface
about the maximum. 
In the approximation that the posterior distribution for the parameters is a
multivariate Gaussian with mean $\vmu\equiv\expec{\x}$ and covariance matrix
$\C\equiv\expec{\x\x^t}-\vmu\vmu^t$, 
its elements are given by \cite{Fisher,Tegmark,VS96,Jungman}
\begin{align}
\label{eq:fish}
F_{ij} = \frac{1}{2}{\rm Tr}
\left[\C^{-1}{\partial\C\over\partial\theta_i}\C^{-1}{\partial\C\over\partial\theta_j}\right]
+{\partial\vmu\over\partial\theta_i}^t\C^{-1}{\partial\vmu\over\partial\theta_j}.
\end{align}
where $\x$ is a N-dimensional vector representing the data set,
whose components $x_i$ are the 
fluctuations in the galaxy cluster density relative to the mean 
in $N$ disjoint cells that cover the three-dimensional 
survey volume in a fine grid. The $\{\theta_i\}$ denote
the cosmological parameters within the assumed fiducial cosmology.
{In the limit where the survey volume is much larger than the scale of 
any features in the observed galaxy cluster power spectrum,
it has been shown \cite{Tegmark97}
that it is possible to redefine $x_n$ in Eq.~(\ref{eq:fish}) to be not
the density fluctuation 
in the $n^{th}$ spatial volume element, but the
average power measured with the FKP method \cite{FKP} in a thin shell 
of radius $k_n$ in Fourier space.

In order to explore the cosmological parameter
constraints from a given galaxy cluster survey, we need to specify
the measurement uncertainties of the cluster power spectrum. 
Therefore, the statistical error on the measurement of the galaxy cluster
power spectrum $P_{\rm c}(k)$ at a given wave-number bin is \cite{FKP}
\begin{equation}
\left[\frac{\Delta P_{\rm c}}{P_{\rm c}}\right]^2=
\frac{2(2\pi)^2 }{\Vsur k^2\Delta k\Delta \mu}
\left[1+\frac1{n_{\rm c}P_{\rm c}}\right]^2,
\label{eqn:pkerror}
\end{equation}
where $n_{\rm c}$ is the mean number density of galaxy clusters given
by Eq.~(\ref{g_final}), 
$\Vsur$ is the comoving survey volume of the galaxy cluster survey, and $\mu$
is the cosine of the angle between ${\bf{k}}$ and the line-of-sight
direction $\mu = \vec{k}\cdot \hat{r}/k$. 

Actually, the \emph{observed} galaxy cluster power spectrum can be different
from the \emph{true} spectrum, and it can be reconstructed
assuming a reference cosmology (which we consider to be our fiducial
cosmology) as (e.g. \cite{SE03})
\begin{align}
P_{\rm obs}(k_{{\rm ref}\perp},k_{{\rm ref}\parallel},z)
=\frac {\DA _{\rm ref} ^2 \hz}{\DA ^2 \hz _{\rm ref}} P_{\rm
  c}(k_{{\rm ref}\perp},k_{{\rm ref}\parallel},z) +P_{\rm shot}\,,
\label{eq:Pobs}
\end{align}
where
\begin{align}
P_{\rm c}(k_{{\rm ref}\perp},k_{{\rm
    ref}\parallel},z)=b_{\mathrm{e}}^2(z)\left[1+\beta(z,k) 
\frac{k_{{\rm ref}\parallel}^2}{k_{{\rm ref}\perp}^2+k_{{\rm
      ref}\parallel}^2}\right]^2\times P_{{\rm L}}(k,z)\,.
\label{eq:Pg}
\end{align}
In Eq.~(\ref{eq:Pobs}), $H(z)$ and $D_A(z)$ are the Hubble parameter
and the angular diameter distance, respectively, and the prefactor 
$(\DA _{\rm ref} ^2 \hz)/(\DA ^2 \hz _{\rm ref})$ encapsulates the
geometrical distortions due to the  Alcock-Paczynski
effect \cite{SE03,9605017,Marulli_etal_2011}.  Their values in the reference cosmology are
distinguished by the subscript `ref', while those in the true cosmology have no
subscript. $k_\perp$ and $k_\parallel$ are the wave-numbers across and along
the line of sight in the true cosmology, and they are related to the 
wave-numbers calculated assuming the reference
cosmology by
$k_{{\rm ref}\perp} = k_\perp D_A(z)/D_A(z)_{\rm ref}$ and
$k_{{\rm ref}\parallel} = k_\parallel H(z)_{\rm ref}/H(z)$. 
$P_\mathrm{shot}$ is the unknown white shot noise that 
remains even after the conventional shot noise of inverse number
density has been  
subtracted \cite{SE03}, and which could arise from clustering bias even
on large scales due to local bias \cite{Seljak00}.
In Eq.~(\ref{eq:Pg}), $b_{\mathrm{e}}(z)$ is the \emph{effective bias}
of Eq.~(\ref{eqn:eb}) between galaxy clusters and 
matter density distributions, and
$\beta(z,k)=f_g(z,k)/b_{\mathrm{e}}(z)$ is the linear  
redshift-space distortion parameter \cite{Kaiser1987}, which in the presence 
of massive neutrinos depends on both redshift and
wave-numbers, since in this case the linear growth rate $f_g(z,k)$ 
is \emph{scale dependent} even at the linear level. We estimate
$f_g(z,k)$ using the 
fitting formula of Ref.~\cite{0709.0253}. For the linear matter power spectrum 
$P_{{\rm L}}(k,z)$, we can encapsulate the effect of
massive neutrino free-streaming into a \emph{redshift dependent}
total matter linear transfer function $T(k,z)$
\cite{0512374,0606533,9710252}, so that  
$P_{{\rm L}}(k,z)$ in Eq.~(\ref{eq:Pobs}) takes the form
\begin{eqnarray}
P_{{\rm L}}(k,z)=\frac{8\pi^2c^4k_0A_s}{25
  H_0^4\Omega_{m}^2} T^2(k,z) \left [\frac{G(z)}{G(z=0)}\right]^2
\left(\frac{k}{k_0}\right)^{n_s}e^{-k^2\mu^2\sigma_r^2},
\label{eq:Pm}
\end{eqnarray}
where $G(z)$ is the usual \emph{scale independent} linear
growth factor in the absence of massive neutrino free-streaming,
i.e. for $k\rightarrow 0$  (see Eq.~(25)
in Ref.~\cite{9710252}), whose fiducial value in each redshift bin is computed 
through numerical integration of the differential equations governing the growth
of linear perturbations in the presence of dark energy
\cite{astro-ph/0305286}. 
The \emph{redshift-dependent}
linear transfer function $T(k,z)$ depends on matter, baryon and massive neutrino
densities (neglecting dark energy at early times), 
and is computed in each redshift bin using
CAMB\footnote{http://camb.info/} \cite{CAMB}.

In Eq.~(\ref{eq:Pm}) we have added the damping factor
$e^{-k^2\mu^2\sigma_r^2}$,
due to redshift uncertainties, where $\sigma_r=(\partial r/\partial
z)\sigma_z$, $r(z)$ being the comoving
distance \cite{SE03,0904.2218}. Here we adopt $\sigma_z=0.001
(1+z)$, for $0.2<z<1$, and $\sigma_z=0.03(1+z)$ for $1<z<2$ \cite{RedBook}, 
since we assume that photometrically selected clusters will 
be confirmed also spectroscopically for low redshifts.

The power spectrum of primordial curvature
perturbations, $P_{\cal R}(k)$, is
\begin{equation}
\Delta^2_{\cal R}(k) \equiv \frac{k^3P_{\cal R}(k)}{2\pi^2}
= A_s\left(\frac{k}{k_0}\right)^{n_s},
\label{eq:pR}
\end{equation}
where $k_0=0.002$/Mpc, $A_s=2.525\times
10^{-9}$ is the dimensionless amplitude of the primordial curvature
perturbations 
evaluated at a pivot scale $k_0$, and $n_s$ is the scalar spectral
index \cite{Larson11}.

Finally, to minimize nonlinear effects, in our Fisher matrix analysis
we restrict wave-numbers to the  
quasi-linear regime, so that, for $z<1$, $\kmax$ is given
by requiring that the variance of matter fluctuations in a sphere of
radius $R$ is $\sigma^2(R)=0.25$ for $R=\pi/(2\kmax)$. This gives 
$\kmax\simeq 0.1 \ihMpc$ at $z=0$ and $\kmax \simeq 0.2 \ihMpc$
at $z=1$, well within the quasi-linear regime. In addition,
for $1<z<2$, we choose $\kmax=0.03 \ihMpc$. Finally, in all the
calculations we impose $k_{\rm min}= 10^{-4} h$/Mpc.

\section{The \emph{Planck} Fisher matrix}
\label{sec:Planck}
In this work we use the \emph{Planck} mission parameter
constraints as CMB priors, by estimating the cosmological parameter errors
via measurements of the temperature and polarization power
spectra. 
As CMB anisotropies, with the exception of the integrated
Sachs-Wolfe effect, are not able to constrain the equation of state of
dark energy $(w_0,w_a)$\footnote{On the contrary, using $(w_0,w_a)$ as
model parameters to compute the CMB Fisher matrix could artificially
break existing degeneracies.}, we follow the prescription laid out by
DETF \cite{DETF,rassat08}. 

We do not include any B-mode in our
forecasts and assume no tensor mode contribution
to the power spectra. 
We use the 100 GHz, 143 GHz, 
and 217 GHz channels as science channels. These channels
have a beam of $\theta_{\rm fwhm}=9.5'$, $\theta_{\rm fwhm}=7.1'$, and
$\theta_{\rm fwhm}=5'$, respectively, 
and sensitivities of $\sigma_T= 2.5 \mu K/K$, $\sigma_T= 2.2 \mu K/K$,
$\sigma_T= 4.8 \mu K/K$ for temperature, and $\sigma_P = 4\mu K/K$,
$\sigma_P = 4.2\mu K/K$, $\sigma_P = 9.8\mu K/K$ for polarization,
respectively.  
We take $f_{\rm sky} = 0.80$ as the sky fraction in order to account for
galactic obstruction, and use a minimum $\ell$-mode $\ell_{\rm
  min}=30$ in order to avoid problems with polarization foregrounds
and not to include information from the late
Integrated Sachs-Wolfe effect, which depends on the specific
dark energy model. 
We discard temperature and polarization data at
$\ell > 2000$  to reduce sensitivity to contributions from patchy
reionisation and point source contamination (see \cite{DETF} and
references therein).

We assume a fiducial cosmology with an anti-correlated isocurvature
contribution, varying dark energy and curvature. 
Therefore we choose the
following set of parameters to describe the temperature and
polarization power spectra 
$\boldsymbol{\theta}= (\omega_m, \omega_b, M_\nu,
100\times \theta_S,\log (10^{10} A_s), n_S, w_0,w_a)$, where $\theta_S$ is the
angular size of the sound horizon at last scattering and $w_0$ and
$w_a$ are the dark energy parameters according to the CPL
parametrization of the dark energy equation of state $w(z) = w_0 + w_a\,z/(1+z)$.

The Fisher matrix for CMB power spectrum is given by \cite{Zalda:1997,Zalda:1997b}:
\begin{equation}
  F_{ij}^{\rm CMB}=\sum_{l}\sum_{X,Y}\frac{\partial
    C_{X,l}}{\partial\theta_{i}}\mathrm{COV^{-1}_{XY}}\frac{\partial
    C_{Y,l}}{\partial\theta_{j}},
  \label{eqn:cmbfisher}
\end{equation}
where $\theta_i$ are the parameters to constrain, 
$C_{X,l}$ is the harmonic power spectrum for the
temperature-temperature ($X\equiv TT$), 
temperature-E-polarization ($X\equiv TE$) and the 
E-polarization-E-polarization ($X\equiv EE$) power spectrum. The covariance
$\rm{COV}^{-1}_{XY}$ of the errors for the various power spectra is
given by the fourth moment of the distribution, 
which under Gaussian assumptions is entirely given in terms of the $C_{X,l}$ with 
\begin{eqnarray}
{\rm COV}_{T,T} & = & f_\ell\left(C_{T,l}+W_T^{-1}B_l^{-2}\right)^2 \\
{\rm COV}_{E,E} & = & f_\ell\left(C_{E,l}+W_P^{-1}B_l^{-2}\right)^2  \\
{\rm COV}_{TE,TE} & = &
f_\ell\Big[C_{TE,l}^2+\left(C_{T,l}+W_T^{-1}B_l^{-2}\right)\left(C_{E,l}+W_P^{-1}B_l^{-2}\right)\Big]\\ 
{\rm COV}_{T,E} & = & f_\ell C_{TE,l}^2  \\
{\rm COV}_{T,TE} & = & f_\ell C_{TE,l}\left(C_{T,l}+W_T^{-1}B_l^{-2}\right) \\
{\rm COV}_{E,TE} & = & f_\ell C_{TE,l}\left(C_{E,l}+W_P^{-1}B_l^{-2}\right)\; ,
\end{eqnarray}
where $f_\ell = 2/((2\ell+1)f_{\rm sky})$,
$W_{T,P}=\sum_c W^c_{T,P}$, $W^c_{T,P}=(\sigma^c_{T,P}\theta^c_{\rm fwhm})^{-2}$  
being the weight per solid angle for temperature and polarization respectively, 
with a 1--$\sigma$ sensitivity per pixel of $\sigma^c_{T,P}$ and a beam
of $\theta^c_{\rm fwhm}$ extent, for each frequency channel $c$. 
The beam window function is given in terms of the full width half
maximum (fwhm) beam width by 
$B_{\ell}^2 =\sum_c (B^c_{\ell})^2 W^c_{T,P}/W_{T,P}$, where 
$(B^c_\ell)^2= \exp\left(-\ell(\ell+1)/(l^c_s)^2\right)$, 
$l^c_s=(\theta^c_{\rm fwhm})^{-1}\sqrt(8\ln2)$ 
and $f_{\rm sky}$ is the sky fraction \cite{9702100}. 

We then calculate the \emph{Planck} CMB Fisher matrix with the help of the
publicly available CAMB code \cite{CAMB}. Finally, we transform 
the \emph{Planck} Fisher matrix for the DETF parameter set to the
final parameter sets ${\bf q}$ considered in this work (see \S~\ref{Fishers}), 
using the transformation
\begin{equation}
F_{\alpha \beta}^{\rm CMB}= \sum_{ij} \frac{\partial
  \theta_i}{\partial q_{\alpha}}\,                       
F_{ij}^{\rm CMB}\, \frac{\partial \theta_j}{\partial q_{\beta}}~.
\end{equation}

\end{document}